\documentclass{article}
\usepackage[numbers]{natbib}
\usepackage[preprint]{neurips_2024}

\usepackage{amssymb}
\usepackage{amsmath}
\usepackage{graphicx}
\usepackage{multicol,multirow}
\usepackage{amsmath,amssymb,amsfonts}
\usepackage{mathrsfs}
\usepackage{amsthm}
\usepackage{rotating}
\usepackage{appendix}
\usepackage{url}
\usepackage{hyperref}
\usepackage[numbers]{natbib}
\usepackage{ifpdf}
\usepackage[T1]{fontenc}
\usepackage{newtxtext}
\usepackage{newtxmath}
\usepackage{textcomp}
\usepackage{xcolor}
\usepackage{lipsum}
\usepackage{pifont}
\usepackage{subcaption}
\usepackage{adjustbox}

\theoremstyle{definition}

\numberwithin{equation}{section}
\usepackage{array}
\makeatletter
\renewcommand{\@noticestring}{Published in \textit{Energy and Buildings}, 2026.}
\makeatother

\title{Conditional Distribution Estimation of Building Characteristics with Diffusion Models for Urban Energy Modeling}

\author{%
  Saumya Sinha$^{1}$\thanks{Corresponding author: saumya.sinha@nlr.gov}, 
  Alexandre Cortiella$^{2}$, 
  Rawad El Kontar$^{1}$, \\
  \textbf{Andrew Glaws$^{1}$, Ryan King$^{1}$, and Patrick Emami$^{1}$} \\
  $^{1}$National Laboratory of the Rockies, $^{2}$Advanced Space
}

\begin{document}

\maketitle
\begin{abstract}
Understanding current energy consumption behavior in communities is critical for informing future energy use decisions and enabling efficient energy management. Urban energy models, which are used to simulate these energy use patterns, require large datasets with detailed building characteristics 
for accurate outcomes. However, such detailed characteristics at the individual building level are often unknown and costly to acquire, or unavailable.
 Through this work, we propose using a generative modeling approach to generate realistic building attributes to fill in the data gaps and finally provide complete characteristics as inputs to energy models. Our model learns complex, building-level patterns from training on a large-scale residential building stock model containing 2.2 million buildings. We employ a tabular diffusion-based framework that is designed to handle heterogeneous (discrete and continuous) features in tabular building data, such as occupancy, floor area, heating, cooling, and other equipment details. We develop a capability for conditional diffusion, enabling the imputation of missing building characteristics conditioned on known attributes. We conduct a comprehensive validation of our conditional diffusion model, firstly by comparing the generated conditional distributions against the underlying data distribution, and secondly, by performing a case study for a Baltimore residential region, showing the practical utility of our approach. Our work is one of the first to demonstrate the potential of generative modeling to accelerate building energy modeling workflows.
\end{abstract}






\section{Introduction}
\label{intro}

Buildings are responsible for significant energy consumption, with the residential sector accounting for 16\% of total energy and 55\% of building energy usage in the U.S.~\cite{EIA}. It is crucial to identify the major factors contributing to the energy consumption, as this knowledge enables more accurate forecasting and therefore informs decisions for improving energy efficiency~\cite{gonzalez2021cross,EPA}. 
The International Energy Agency (IEA) projects that buildings could reduce energy consumption by 40\% by 2040~\cite{efficiency2019buildings,el5264131ai}. Achieving such progress in energy conservation, however, requires complete knowledge of energy use behavior and detailed characteristics at the building-level resolution. Such datasets are essential for studying fine-grained energy use patterns. They are also important for optimizing conservation techniques for different kinds of communities and adopting building-specific technological upgrades~\cite{el2024open,el5264131ai}.

Urban Building Energy Models (UBEMs) simulate energy use at different scales, and tools like URBANopt~\cite{el2020urbanopt,polly2016zero} operate at a district level. Their modeling accuracy depends on localized building characteristics and geographical information. Precise details on building footprints, occupant behavior, and socioeconomic factors are needed for these simulations, but such data are not always available or cannot be inferred reliably from sources like satellite imagery~\cite{biljecki2023quality}. This creates a data gap, and filling this with, e.g, the default prototype specifications used in URBANopt~\cite{IECC2021,ASHRAE2024} is insufficient, as they are not representative of the diversity in building and occupant energy use~\cite{el2024open,el5264131ai}. Moreover, large-scale synthetic datasets such as the ResStock~\cite{wilson2017resstock} dataset, which is statistically representative of the entire residential building stock in the U.S., model energy use at a Public Use Microdata
Area (PUMA)-level, thus providing regional information but lacking the required accurate details at building-level~\cite{el2024open,el5264131ai}.  

Deep learning models, specifically generative models, offer a promising solution by generating synthetic data to fill in the data gaps we observe in the building stock datasets. Among these, recent diffusion models
~\cite{sohl2015deep,ho2020denoising,song2019generative,rombach2022high} have shown better potential than existing generative models, such as Generative Adversarial Networks (GANs), in domains like computer vision and natural language processing (NLP)~\cite{dhariwal2021diffusion,nichol2021improved,kotelnikov2023tabddpm}. However, generative approaches, and diffusion models in particular, have had limited exploration in the area of building energy modeling. Given that the datasets in this domain are primarily tabular in nature, our modeling framework is based on 
TabDDPM~\cite{kotelnikov2023tabddpm}, which is a Denoising Diffusion Probabilistic Model (DDPM)~\cite{sohl2015deep,ho2020denoising}, specifically developed for tabular datasets. While learning to model such datasets and generating high-quality tabular synthetic data is in demand, it is a challenging task, as it involves working with heterogeneous features. Tabular data is inherently composed of mixed-type feature distributions, e.g., some feature columns are discrete while others are continuous. Furthermore, it is often difficult to obtain internet-scale massive tabular datasets that are typically used in training generative models. TabDDPM is designed to overcome these challenges and has been shown to outperform other generative models, like GANs and Variational Autoencoder (VAEs) based models, in handling numerical as well as categorical data across many datasets. 
\begin{figure}[!htbp]
    \centering
    \includegraphics[width=\textwidth]{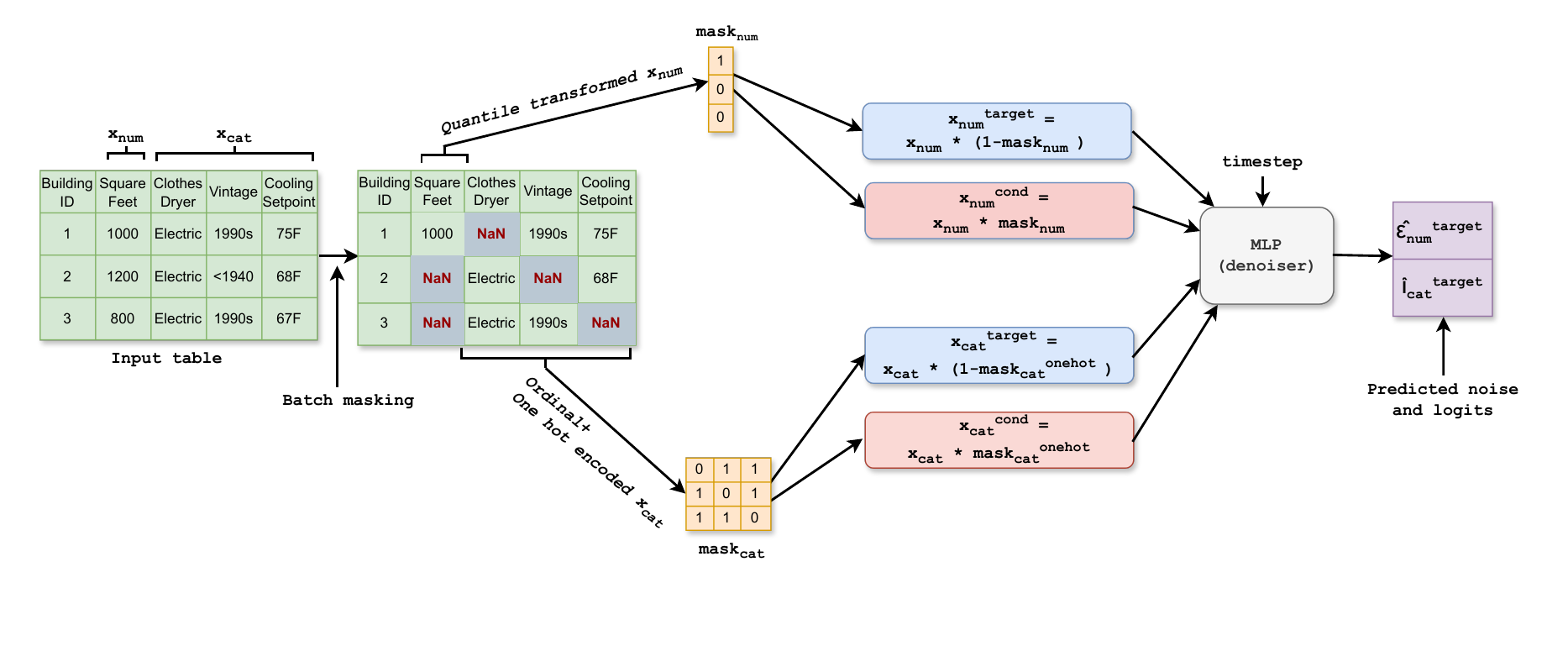}
    \caption{
    Training overview of our conditional generative diffusion model for
a mixed-type tabular dataset. The Conditional TabDDPM model learns to generate the missing (unknown) building characteristics conditioned on observed building attributes. 
We show details of the denoiser training pipeline for a single timestep in this figure, where the known (\textit{condition}) features are used to guide the denoising of the unknown (\textit{target}) features. A training batch of building characteristics consists of numerical ($x_{num}$) and categorical features ($x_{cat}$), which are encoded with their respective transforms. 
    A random subset of features is then masked, and these masks ($mask_{num}$ and $mask_{cat}$) are used to create the \textit{target} and \textit{condition} components. The (noisy) \textit{target} and (observed) \textit{condition} components are then concatenated and passed as inputs to the MLP denoiser model, along with the current diffusion timestep (embedded as in~\cite{kotelnikov2023tabddpm}).  The MLP is trained to predict the noise for numerical and logits for categorical \textit{target} variables, corresponding to this timestep.
    }
    \label{fig:overall_framework}
\end{figure}

The main focus of our work is on adapting the TabDDPM model to perform conditional generation, to impute or fill in the missing building characteristics conditioned on the building location, its energy use, and other observed building details. This yields a single, foundational diffusion model capable of generating realistic building characteristics and advancing our ultimate goal of filling in the unknown characteristics to create a complete, fine-grained building-level dataset to use as input to building energy modeling workflows. 
To enable this, we train our conditional diffusion model on the ResStock dataset. ResStock is a tabular dataset, 
where each row corresponds to an individual building (sampled from their PUMA-level distributions), and columns 
consist of the building's characteristics, such as square footage, HVAC (Heating, Ventilation, and Air Conditioning) systems, occupant information, various equipment details, and their usage levels. Our generative modeling framework learns these complex, mixed-type multivariate conditional distributions and captures intricate dependencies between all detailed attributes.  

We perform an extensive evaluation of our Conditional TabDDPM model's generation capabilities. We first compare the generated conditional distributions against the true distributions from the original ResStock dataset.
Furthermore, our evaluation includes a real-world case study focused on a residential area in Baltimore, Maryland. Here, our model's generated values fill in varying levels of unknown characteristics to generate complete building data, which are then passed as inputs to the URBANopt energy model. 
We validate the effectiveness of this approach by comparing URBANopt's resulting energy load profile against the load profiles obtained using a curated (reference) dataset for the community. This case study shows an end-to-end application of our conditional generative modeling approach. Overall, our encouraging evaluation demonstrates the potential of conditional generation with TabDDPM to accelerate building energy modeling workflows.

To summarize, our main contributions are as follows: \\
1) We develop a conditional generation framework using a diffusion-based generative model (TabDDPM) that learns multivariate conditional distributions over complex building characteristics. The model can generate diverse, realistic values for any combination of unknown building characteristics conditioned on observed attributes. Our approach handles heterogeneous (discrete and continuous) features common in ResStock's tabular building data, such as building square footage, total electricity consumption, occupancy, and HVAC details. \\
2) We showcase the use of our Conditional TabDDPM model in a real-world building energy modeling workflow. In this case study, we employ our model to generate multiple plausible scenarios for unknown characteristics and fill in these gaps in the building characteristics dataset for a residential region in Baltimore. This creates complete inputs for the URBANopt energy model, and we validate its energy load profile outputs. 

\section{Related Work}
\label{related_work}
\subsection{Deep Learning in Energy Systems and Buildings application} 
\label{related_work_ml_for_buildings_application}
Data-driven approaches, such as Machine Learning (ML) methods, promise better efficiency over expensive numerical simulations for energy systems modeling~\cite{carter2023advanced}. Deep learning models have been explored for predicting building energy consumption~\cite{lu2022building,vazquez2019deep,amasyali2018review,zhang2021review,emami2024syscaps,emami2023buildingsbench}. As building energy use prediction is typically a time series forecasting or a regression problem, reviews such as~\cite{lu2022building,zhang2021review} have discussed the suitability of ML models (e.g., support vector machines, neural networks, recurrent neural networks) for this task. More recently, works like BuildingsBench~\cite{emami2023buildingsbench} 
introduce a large-scale dataset of simulated buildings,
and use it to pre-train transformer models, benchmarking their short-term load forecasting performance on real residential and commercial buildings. 
Concurrently, models such as Graph neural networks are used to predict building characteristics~\cite{lei2024predicting}, to complete building information in geospatial datasets, and facilitate urban studies. With the advance of deep learning in these applications, there is an increasing interest in using deep generative models~\cite{wu2022generative}. In recent reviews~\cite{zhang2025deep,glaws2025designing}, the authors enlist important application areas in the overall field of energy systems where generative models have contributed, while also pointing out their limited exploration in the specific area of building energy management.

Our work is related to recent studies by El Kontar et al.~\cite{el2024open,el5264131ai}, which focus on improving localized (district-level) building energy modeling using ML frameworks to fill data gaps in building characteristics data and reverse engineer the bottom-up modeling framework~\cite{kavgic2010review}. They achieve this by using deep neural networks to learn the relationship between known and unknown building attributes and predict missing characteristics, generating complete inputs for the URBANopt energy model. Their work integrates data with multiple modalities, including time series inputs from sources such as ResStock~\cite{wilson2017resstock}, OpenStreetMap, and Zillow/Redfin~\cite{el5264131ai}. The modeling framework also provides scenario generation capabilities for various efficiency targets chosen by the user via an injection approach.

Our proposed work differs fundamentally by developing a generative modeling framework instead of a deterministic
one. This is a more challenging and powerful approach, as our generative model learns the entire underlying conditional data distribution over multiple building characteristics, rather than 
a direct point
mapping. We train a single, unified diffusion model with conditional generation capabilities that can be sampled to generate diverse plausible values for any number of missing building characteristics conditioned on the observed attributes. 
Our model is trained once on the comprehensive ResStock dataset, which represents the entire U.S. residential building stock. This enables it to be applied across diverse geographical locations, as opposed to training a separate neural network model for each district (\cite{el2024open,el5264131ai}).

\subsection{Deep Learning for Tabular Data Imputation}
Tabular datasets are ubiquitous in many domains. Classical statistical methods like MICE~\cite{van2011mice} provide a standard baseline for tabular data imputation by iteratively estimating missing values, but they scale poorly to large datasets and struggle to capture complex, non-linear feature interactions~\cite{shah2014comparison}. Many deep learning approaches for tabular data overcome these limitations, and treat imputation as a deterministic task, mapping available inputs to a single point estimate for the missing value~\cite{borisov2022deep}. Recent models utilizing masked autoencoding frameworks~\cite{du2023remasker}, contrastive learning~\cite{kowsar2025deepifsac}, and implicit neural representations~\cite{ochs2025tabinr} have improved prediction accuracy on incomplete datasets. Concurrently, modern architectures like TabM~\cite{gorishniy2025tabm} have refined how neural networks process tabular structures compared to traditional tree-based models (e.g. gradient-boosted decision tress~\cite{chen2016xgboost}) which have historically dominated this domain~\cite{shwartz2022tabular}. However, these methods learn a point mapping to an expected value rather than capturing the underlying distribution of the missing data, highlighting the need for deep generative models that learn the joint distribution over features, as discussed below.

\subsubsection{Generative models for Tabular Data}
Developing generative models for tabular data has many important use cases~\cite{shi2025tabdiff,kotelnikov2023tabddpm}. High-quality synthetic tabular generation could be used for data augmentation and is in demand for its privacy-preserving benefits. 
Generative models also provide capabilities for missing value imputation by framing it as a conditional generation task~\cite{yoon2018gain,yoon2020gamin}, which is the primary focus of our study. However, tabular data presents a challenge as it is often comprised of heterogeneous or mixed-type features, requiring the generative approach to jointly learn distributions over both discrete and continuous variables.

GANs~\cite{xu2019modeling,zhao2021ctab,zhao2024ctab} and VAEs~\cite{xu2019modeling,liu2023goggle} have been adapted for tabular data generation, but with suboptimal generation quality. TabMT~\cite{gulati2023tabmt} uses a masked transformer, while a recent work~\cite{anshelevich2025synthetic} combines GAN and VAE for the tabular generation task. Recently, diffusion models have shown increasing promise in tabular data synthesis~\cite{zheng2022diffusion,zhang2024mixedtype,kotelnikov2023tabddpm,shi2025tabdiff,lee2023codi,villaizan2024diffusion,kim2023stasy,hudovernik2025reldiff}. Some works~\cite{zheng2022diffusion,zhang2024mixedtype} encode mixed features in a single latent continuous space and apply Gaussian diffusion, thus simplifying the diffusion model training on heterogeneous features, but having an additional encoding overhead. Other models like TabDDPM~\cite{kotelnikov2023tabddpm} and CoDi~\cite{lee2023codi} use discrete-time diffusion processes, separately for numerical (continuous) and categorical (discrete) features. Graph-based diffusion models are used for generation on relational data~\cite{hudovernik2025reldiff}, which consist of complex interlinked tables. Furthermore, TabDiff~\cite{shi2025tabdiff} uses a joint continuous-time diffusion framework, tackling the feature heterogeneity by using feature-wise learnable diffusion processes. 

Some of these diffusion frameworks have been specifically developed to perform tabular data imputation. Zheng et al.~\cite{zheng2022diffusion} adapt the CSDI~\cite{tashiro2021csdi} diffusion model, originally developed for missing value imputation in time-series data, to handle numerical and categorical variables in tabular data using various embedding techniques. Work in~\cite{villaizan2024diffusion} proposes architectural changes such as using a novel conditioning attention mechanism to capture the complex relationship between known and unknown tabular features.
The TabDiff model~\cite{shi2025tabdiff} also performs conditional generation for missing value imputation using classifier-free guidance~\cite{ho2022classifier}. However, the classifier-free guidance framework is limited as it needs to train a new, small, specialized unconditional model for every unique combination of columns to be imputed. 

Our work builds on Kotelnikov et al.'s TabDDPM model~\cite{kotelnikov2023tabddpm}. While the latest diffusion models, such as TabDiff~\cite{shi2025tabdiff}, have been shown to outperform TabDDPM on some benchmarks, the main goal of this study is to adapt a simpler, discrete-time diffusion framework of TabDDPM to perform robust conditional generation for building characteristics, providing an accessible but effective tool for the building energy modeling community. We describe our modeling framework in detail in the next section.

\section{Methodology}
\label{methodology}
In this section, we describe our modeling framework: a conditional generative diffusion model for a mixed-type tabular dataset. We refer to our model as Conditional TabDDPM (Figure~\ref{fig:overall_framework}), as it builds upon the TabDPPM~\cite{kotelnikov2023tabddpm} network. 
\subsection{Background}
\label{background}
The TabDDPM model is a Denoising Diffusion Probabilistic Model (DDPM)~\cite{sohl2015deep,ho2020denoising}, specifically designed for tabular data synthesis. TabDDPM employs two separate diffusion models to handle mixed-type data: a Gaussian diffusion model for numerical features and a Multinomial diffusion model for categorical features~\cite{kotelnikov2023tabddpm}. We will briefly introduce the general diffusion framework along with the Gaussian and Multinomial diffusion models, before describing the specifics of our Conditional TabDDPM model. 

Diffusion models are a generative modeling framework that learn the data distribution through a forward and reverse Markov process. The forward process  \(q(x_{1:T}\mid x_0)=\prod_{t=1}^{T} q(x_t\mid x_{t-1})\) iteratively adds noise to an initial sample $x_0\!\sim q(x_0)$ over \( T\) timesteps. At each step, noise is sampled from a predefined distribution $q(x_t\mid x_{t-1})$ with variances $\{\beta_1,\ldots,\beta_T\}$. The reverse process, \(p(x_{0:T})=\prod_{t=1}^{T} p(x_{t-1}\mid x_t)\) denoises $x_T\!\sim q(x_T)$ at every timestep, allowing new data samples to be generated from $q(x_0)$. This reverse process is approximated by a neural network (parametrized by \(\theta\)), trained on the data by optimizing the following variational lower bound~\cite{kotelnikov2023tabddpm}: 
\begin{equation}
\label{eq:elbo}
\begin{split}
\log q(x_0)
\;\ge\;
\mathbb{E}_{q(x_0)}\!\left[
\underbrace{\log p_{\theta}(x_0\mid x_1)}_{L_0}
\right.
& - \underbrace{\operatorname{KL}\!\big(q(x_T\mid x_0)\,\|\,q(x_T)\big)}_{L_T} \\
& \left. - \sum_{t=2}^{T}
\underbrace{\operatorname{KL}\!\big(q(x_{t-1}\mid x_t, x_0)\,\|\,p_{\theta}(x_{t-1}\mid x_t)\big)}_{L_t}
\right].
\end{split}
\tag{1}
\end{equation}

Gaussian diffusion operates in continuous space ($x_t \in \mathbb{R}^d$)~\cite{kotelnikov2023tabddpm}. The forward process is defined by a Gaussian distribution \( q(x_t \mid x_{t-1}) := \mathcal{N}\left(x_t;\, \sqrt{1 - \beta_t} \, x_{t-1},\, \beta_t I\right) \) and the reverse process is characterized as \(p_{\theta}(x_{t-1}\mid x_t):= \mathcal{N}\!\big(x_{t-1};\, \mu_{\theta}(x_t,t),\, \Sigma_{\theta}(x_t,t)\big)\). Following Ho et al.~\cite{ho2020denoising}, we use diagonal \(\Sigma_{\theta}(x_t, t)\) with constant $\sigma_t$ and:
\begin{equation}
\label{eq:mu-param}
\mu_{\theta}(x_t,t)
= \frac{1}{\sqrt{\alpha_t}}
\left(
x_t - \frac{\beta_t}{\sqrt{1-\bar{\alpha}_t}}\,\epsilon_{\theta}(x_t,t)
\right)
\qquad
t \in \{1,\ldots,T\},
\tag{2}
\end{equation}
where \(\alpha_t := 1-\beta_t\), \(\bar{\alpha}_t := \prod_{i=1}^{t}\alpha_i\). \(\epsilon_{\theta}(x_t,t)\) is the predicted noise component in the noisy $x_t$. This simplifies the neural network training to minimize the mean-squared error (MSE) between the ground truth noise  \(\epsilon\) in \(x_t\) and the predicted noise \(\epsilon_{\theta}(x_t,t)\) over all timesteps~\cite{kotelnikov2023tabddpm}.
\begin{equation}
\label{eq:gaussian-loss}
L^{\text{simple}}
= \mathbb{E}_{x_0,\epsilon,t}\!\left[\;\bigl\lVert \epsilon - \epsilon_{\theta}(x_t,t)\bigr\rVert_2^{2}\right]
\tag{3}
\end{equation}

Multinomial diffusion handles categorical data with one-hot encoded categorical inputs as \( x_t \in \{0,1\}^K \) (\( K \) classes)~\cite{kotelnikov2023tabddpm}. Following~\cite{hoogeboom2021argmax}, the forward process mixes the data with uniform noise
over all classes, defined as the categorical distribution \( q(x_t \mid x_{t-1}) = \mathrm{Cat}\big(x_t;\,(1 - \beta_t) x_{t-1} + \tfrac{\beta_t}{K}\big) \) and \( q(x_t \mid x_0) = \mathrm{Cat} \left(x_t; \bar{\alpha}_t x_0 + \frac{1 - \bar{\alpha}_t}{K} \right) \). From these two equations, the posterior can be computed as \(q(x_{t-1} \mid x_t, x_0) = \mathrm{Cat}\!\left(x_{t-1}; \frac{\pi}{\sum_{k=1}^K \pi_k}\right),\quad\text{where}\)
\(\pi = \big[\alpha_t x_t + (1-\alpha_t)/K\big] \odot \big[\bar{\alpha}_{t-1}x_0 + (1-\bar{\alpha}_{t-1})/K\big]\). The reverse process \( p(x_{t-1} \mid x_t) \) is paramterized using \( q(x_{t-1} \mid x_t, \hat{x}_0) \), where \(\hat{x}_0 = \mu(x_t, t)\) is approximated by a neural network, optimized based on the KL divergence from Equation~\ref{eq:elbo}.

\subsection{Conditional TabDDPM}
\label{condtabddpm}
We adapt the TabDDPM~\cite{kotelnikov2023tabddpm} framework to perform conditional diffusion, where the goal is to generate or impute values for the unobserved (unknown) variables conditioned on a set of observed variables. The observed or unobserved variables can be any combination of the numerical or categorical features. Our Conditional TabDDPM model is designed to handle heterogeneous features, i.e., numerical and categorical data in tabular datasets with rows of the form \( x: [x_{\text{num}}, x_{\text{cat}_1}, \ldots, x_{\text{cat}_C}] \). Here, we have $N_{num}$ numerical and $C$ categorical features, each \( x_{\text{cat}_i} \) has \( K_i \) categories. 
 Each row $x$ is partitioned into \textit{condition} and \textit{target} components, i.e. \( x^{\text{cond}} \) for observed variables and \(x^{\text{target}} \) for unobserved variables (as seen in~\cite{zheng2022diffusion,villaizan2024diffusion}). The model takes in quantile-normalized numerical features and one-hot encoded categorical features. Gaussian diffusion is applied to \(x^{\text{target}}_{num}\), and Multinomial diffusion to \(x^{\text{target}}_{cat}\), both conditioned on \(x^{\text{cond}}\). 


The forward process is applied only to the \textit{target} (unobserved) variables, \(x^{\text{target}} \),
and each categorical feature has a separate forward diffusion process, as its noise is sampled independently~\cite{kotelnikov2023tabddpm}. The reverse diffusion process denoises \( x_{t}^{\text{target}}\) conditioned on \(x^{\text{cond}} \) observed variables. Figure~\ref{fig:overall_framework} provides an overview of the denoiser training pipeline. The reverse process is modeled as \( p(x_{t-1}^{\text{target}} \mid x_t^{\text{target}},\, x^{\text{cond}}) \), where \(x^{\text{cond}} \) remains fixed across all timesteps, as opposed to the unconditional generation modeling \( p(x_{t-1} \mid x_t) \) introduced in~\ref{background}. 
Our denoising network, modeling the reverse diffusion process (Figure~\ref{fig:overall_framework}), is a single multi-layered perceptron (MLP) that predicts the noise for numerical and logits for categorical 
variables, corresponding to the timestep. During training, we perform dynamic masking~\cite{villaizan2024diffusion}, which presents the model with a different, randomly selected set of observed variables every time. This improves the robustness of the conditional generation capability. The MLP is trained on a sum of the MSE loss from Equation~\ref{eq:gaussian-loss} for the Gaussian diffusion and the KL-divergence loss, from Equation~\ref{eq:elbo} for the Multinomial diffusion of each of the categorical features. The loss is computed only on the \textit{target} variables~\cite{kotelnikov2023tabddpm,villaizan2024diffusion}:
\begin{equation}
L_t^{condTab} = L_t^{simple} \, (\textit{target}) + \frac{\sum_{i \leq C} L_t^i \, (\textit{target})}{C}
\label{eq:tabddpm_loss}
\tag{4}
\end{equation}


Here, we discuss additional details of the conditional masking process. A binary mask $mask_\text{num}$ and $mask_\text{cat}$ is created for both the numerical and categorical features, with $1$ indicating observed (\textit{condition}) and $0$ indicating unobserved or masked (\textit{target}) variables. Specifically, during training, a random subset of numerical features is chosen, for each tabular row, to create a \(\{0,1\}^{N_{\text{num}}} \) numerical mask, and similarly a categorical mask of size \(\{0,1\}^{\sum_{i=1}^{C} K_i} \) is created for the one-hot encoded categorical features. These masks ($mask_\text{num}$ and $mask_\text{cat}$) are then used to partition the inputs $x$ into \( x^{\text{cond}} \) and \(x^{\text{target}} \) components (see Figure~\ref{fig:overall_framework}). The denoiser MLP model, adapted from TabDDPM~\cite{kotelnikov2023tabddpm}, is modified to accept a concatenated input of the form: 
\( (x^{\text{target}}_{t,num},x^{\text{target}}_{t,cat},x^{\text{cond}}_{num},x^{\text{cond}}_{cat}) \), 
and we only keep the model output values corresponding to the \textit{target} variables. 

During sampling or inference, is it known which variables are observed and which are unobserved. We begin the reverse diffusion process with noise for the \textit{target} variables and iteratively denoise them over a total of $T$ timesteps conditioned on the observed variables. The final outputs are the sampled values for the \(x^{\text{target}} \) variables, which are then postprocessed to reverse the normalization and encoding.
\\
\\
\textbf{Training details: } We train two variants of the Conditional TabDDPM model:
(1) mixed-imputation model: a model trained to impute a mix of both numerical and categorical features, and (2) categorical-only imputation model: a model trained to impute only categorical features while always observing all numerical features. For (2), we only train with the categorical loss. For (1), we use a weighted loss function shown below, modified over~\ref{eq:tabddpm_loss}, where \(\lambda_{\text{num}}\) for the numerical component is fixed to $1$, and \(\lambda_{\text{cat}}\) for the categorical component is a tuned hyperparameter.
\begin{equation}
L_t^{condTab_{weighted}} = \lambda_{\text{num}} \, L_t^{simple} \, (\textit{target}) 
+ \lambda_{\text{cat}} \, \frac{\sum_{i \leq C} L_t^i \, (\textit{target})}{C}
\label{eq:tabddpm_loss_weighted}
\tag{5}
\end{equation}

The numerical features are transformed with the Gaussian quantile transformation~\cite{pedregosa2011scikit}, and the categorical features are ordinal encoded before being one-hot encoded. We use parameters \(f_{\text{mask, num}}\) and \(f_{\text{mask, cat}}\) to control for the maximum proportion of the numerical variables and categorical variables to be masked while training. During each training step, a random positive masking ratio between \((0,f_{\text{mask, num}})\) and \((0,f_{\text{mask, cat}})\) is selected, and the corresponding number of numerical and categorical variables (respectively) is masked or set unobserved for every row in the batch. Notably, while the same number of variables is masked from every row in a batch, the variables themselves are chosen at random. This is the dynamic masking strategy described above. To this end, for (1), we train a single model with  \(f_{\text{mask, num}} = 0.5\) and \(f_{\text{mask, cat}} = 0.2\). For (2), we train two separate models with all numerical features observed and \(f_{\text{mask, cat}}\) set to 0.05 and 0.4, respectively. All Conditional TabDDPM models were trained on a single NVIDIA H100 GPU. The training time for the mixed-imputation model was 4.53 hours, and both the categorical-only variants took less than 3 hours. 
\section{Evaluation}
\label{evaluation}
\textbf{Dataset}: Our study utilizes the ResStock~\cite{wilson2017resstock} dataset (\url{https://resstock.nlr.gov/}). ResStock is a synthetic dataset statistically representative of the entire residential
building stock in the U.S~\cite{emami2023buildingsbench}. This dataset is generated by sampling from probability distributions derived from multiple data sources~\cite{wilson2017resstock}, resulting in an extensive collection of U.S residential buildings at PUMA-level resolution~\cite{el2024open,el5264131ai}. A PUMA varies in area, consists of many geographically close buildings, and contains granular details on building characteristics in the area. We use the 2024.1 release version of the ResStock dataset, which comprises 2.2 million buildings. Each row of this tabular data corresponds to a unique building, and the columns include various building characteristics and energy consumption details. From this list of available features, we select a small subset to work with: 
\\
(a) Numerical (continuous) features: these include the building square footage and total electricity consumption columns. We also add a latitude and a longitude feature for each building using the centroid of the building's PUMA to encode its geospatial information~\cite{emami2023buildingsbench}. This helps to replace the categorical PUMA variable, which has $\sim2.4$k categories. \\
(b) Categorical (discrete) features: these include building attributes such as equipment details (e.g., clothes dryer, dishwasher, refrigerator, HVAC heating/cooling system), equipment usage levels, geometric stories, building age, and occupant information. 

In total, we have 35 features (4 numerical and 31 categorical). We always keep the latitude and longitude columns known or observed in our conditional generation setup. We use a small subset of withheld PUMAs~\cite{emami2023buildingsbench} to create an out-of-distribution (OOD) test dataset for evaluating our generative model's performance in unseen geographical regions. Overall, we split the dataset into train (1,405,212), validation (351,303), test (439,128), and OOD-test set (3,665).
\\
\\
\textbf{Hyperparameter Tuning:} We adopt most of the hyperparameters from the TabDDPM paper~\cite{kotelnikov2023tabddpm}, such as the batch size, diffusion timesteps, training iterations, scheduler, and the MLP dropout. 
We tune other hyperparameters: sweeping learning rate over \{1e-3, 5e-4, 1e-4\} and MLP model architectures over a standard 6-layer network $[256,512,512,512,512,256]$, a wide 6-layer network with double width per layer, and a deep 8-layer network with two hidden layers added to the standard architecture. We also tuned \(\lambda_{\text{cat}}\) (for Equation~\ref{eq:tabddpm_loss_weighted}) by testing both fixing it to 1, and a linear weight decay schedule initialized with 1. To choose the best hyperparameters, we track reconstruction MSE and Accuracy scores for a randomly chosen target set of numerical and categorical features in the validation dataset (reconstruction scores discussed in detail in~\ref{reconstruction_evaluation}). For the Conditional TabDDPM model variant (1), the best performance was obtained with the standard MLP model, with a learning rate of 5e-4 and a decaying \(\lambda_{\text{cat}}\). For variant (2), we chose the same MLP setup but with a learning rate of 1e-4. Finally, we train the best-performing models for both (1) and (2) for 3 different seeds. 

\subsection{Conditional generation evaluation}
\label{conditional_generation_evaluation}
\begin{table}[!htbp]
\centering
\caption{\textbf{Univariate conditional generation results}. The table shows the distance between the true and the generated distribution, for each of the imputable variables, averaged across runs with the three random seeds. We use Wasserstein distance for the two numerical features and JS distance for the 31 categorical features, following~\cite{kotelnikov2023tabddpm, zhao2021ctab,villaizan2024diffusion}. The lower the distances, the better. The
results shown are obtained with the mixed imputation model, which handles both numerical and categorical feature imputation. The last row includes the JS distance for the random baseline, averaged across all variables, added to compare against our model's performance on categorical features.} 
\vspace{2ex}
\label{tab:univatiate_analysis}
\begin{adjustbox}{max width=\columnwidth}
\footnotesize
\begin{tabular}{l r} 
\hline \hline 
\textbf{Feature Name} & \textbf{Distance} \\
\hline 
\multicolumn{2}{l}{\textit{Numerical Features (Wasserstein Distance)}} \\
Sqft (square footage)                    & 0.113 \\ 
Total electricity consumption & 0.116 \\
\hline
\textbf{Average Numerical WD}         & \textbf{0.115} \\
\hline

\multicolumn{2}{l}{\textit{Categorical Features (JS Distance)}} \\
Clothes dryer                         & 0.337 \\
Clothes dryer usage level             & 0.022 \\
Clothes washer                        & 0.309 \\
Clothes washer usage level            & 0.026 \\
Dishwasher                            & 0.092 \\
Dishwasher usage level                & 0.025 \\
Cooling setpoint                      & 0.158 \\
Cooling setpoint has offset           & 0.479 \\
Cooling setpoint offset magnitude     & 0.036 \\
Heating fuel                          & 0.485 \\
Heating setpoint                      & 0.362 \\
Heating setpoint has offset           & 
0.561 \\
heating setpoint offset magnitude     & 0.041 \\
Refrigerator                          & 0.088 \\
Refrigerator usage level              & 0.024 \\
HVAC cooling efficiency               & 0.062 \\
HVAC cooling type                     & 0.351 \\
HVAC has ducts                        & 0.182 \\
HVAC has zonal electric heating       & 0.004 \\
HVAC heating efficiency               & 0.301 \\
HVAC heating type                     & 0.336 \\
HVAC heating type and fuel            & 0.071 \\
HVAC system is faulted                & 0.0 \\
Water heater efficiency               & 0.187 \\
Water heater fuel                     & 0.418 \\
Water heater in unit                  & 0.059 \\
Occupants                             & 0.384 \\
Neighbors                             & 0.033 \\
Geometry stories                      & 0.079 \\
Geometry building type recs (building type)           & 
0.499 \\
Vintage                               & 0.305 \\
\hline
\textbf{Average Categorical JSD}         & \textbf{0.204} \\
\hline 

\textbf{Random baseline average JSD}      & \textbf{0.462} \\
\hline 
\end{tabular}
\end{adjustbox}
\end{table}

We evaluate the quality of the imputed features in the test and OOD-test dataset, generated by our Conditional TabDDPM model, with the following:
\\
\textbf{Metrics:} We evaluate the imputed features by comparing their generated distributions against the true distributions observed in the training data. For a quantitative comparison, we follow prior work~\cite{kotelnikov2023tabddpm, zhao2021ctab,villaizan2024diffusion}, and use the Jensen-Shannon (JS) distance metric to quantify the distance between these distributions for categorical variables, and the Wasserstein distance for numerical distributions. In addition to the quantitative metrics, our analysis includes a qualitative assessment which involves visualizing the generated conditional distribution \( p(x^{\text{gen}} \mid x^{\text{cond}} )\) alongside the true conditional distribution \( p(x^{\text{true}} \mid x^{\text{cond}} )\). 

We include univariate and bivariate conditional generation analysis as a part of this evaluation.
\subsubsection{Univariate analysis} 
For the univariate analysis, we examine each imputed feature ($x_i$) individually, by comparing its generated conditional distribution \( p(x_i^{\text{gen}} \mid x^{\text{cond}} )\) to the corresponding true distribution \( p(x_i^{\text{true}} \mid x^{\text{cond}} )\).
\\
\\
\textbf{Estimating \( p(x_i^{\text{gen}} \mid x^{\text{cond}} )\) and \( p(x_i^{\text{true}} \mid x^{\text{cond}} )\): } 
We mask every variable, $x_i$, in the test dataset, one at a time, systematically evaluating each imputable feature. For the masked variable $x_i$ (the \textit{target} variable in this case), we first identify its known dependencies from the ResStock documentation. 
These are a subset of the rest of the observed variables (\(x^{\text{cond}}\)). A list of all dependencies for each building characteristic can be obtained from ResStock's webpage, e.g, the type of clothes dryer appliance depends on the fuel used by the building's heating system and the building type. We then iterate through every unique combination of these dependency values present in the test data. For every unique set of dependencies, we build a true distribution \( p(x_i^{\text{true}} \mid x^{\text{cond}} )\) and a generated distribution \( p(x_i^{\text{gen}} \mid x^{\text{cond}} )\). The true distribution is constructed by finding all rows in the training set with the matching dependency values and sampling $1000$ rows (if the number of matches exceeds that) to get the \(x_i^{\text{true}}\) values. The focus on matching a dependency subset instead of all of \(x^{\text{cond}}\) variables is motivated by the challenge of finding very few to no training data when doing an exact match with these many columns. Moreover, this domain knowledge is useful as it identifies which variables could potentially be important in informing $x_i$. We construct the generated distribution with an equal number of \(x_i^{\text{gen}}\) synthetic samples, obtained by repeatedly sampling from the conditional TabDDPM model conditioned on the same \(x^{\text{cond}}\), yielding a distribution of plausible values. 

For $x_i$s that have no defined dependencies, the true distribution is created using the values of $x_i$ in $1000$ random rows selected from the training data. We create the generated distribution by imputing $x_i$ in $1000$ randomly selected test data rows, using the rest of the variables in the row as \(x^{\text{cond}}\). 
\\
\begin{figure}[htbp]
    \centering 
    \begin{subfigure}[b]{0.48\textwidth}
        \centering
        \includegraphics[width=\textwidth]{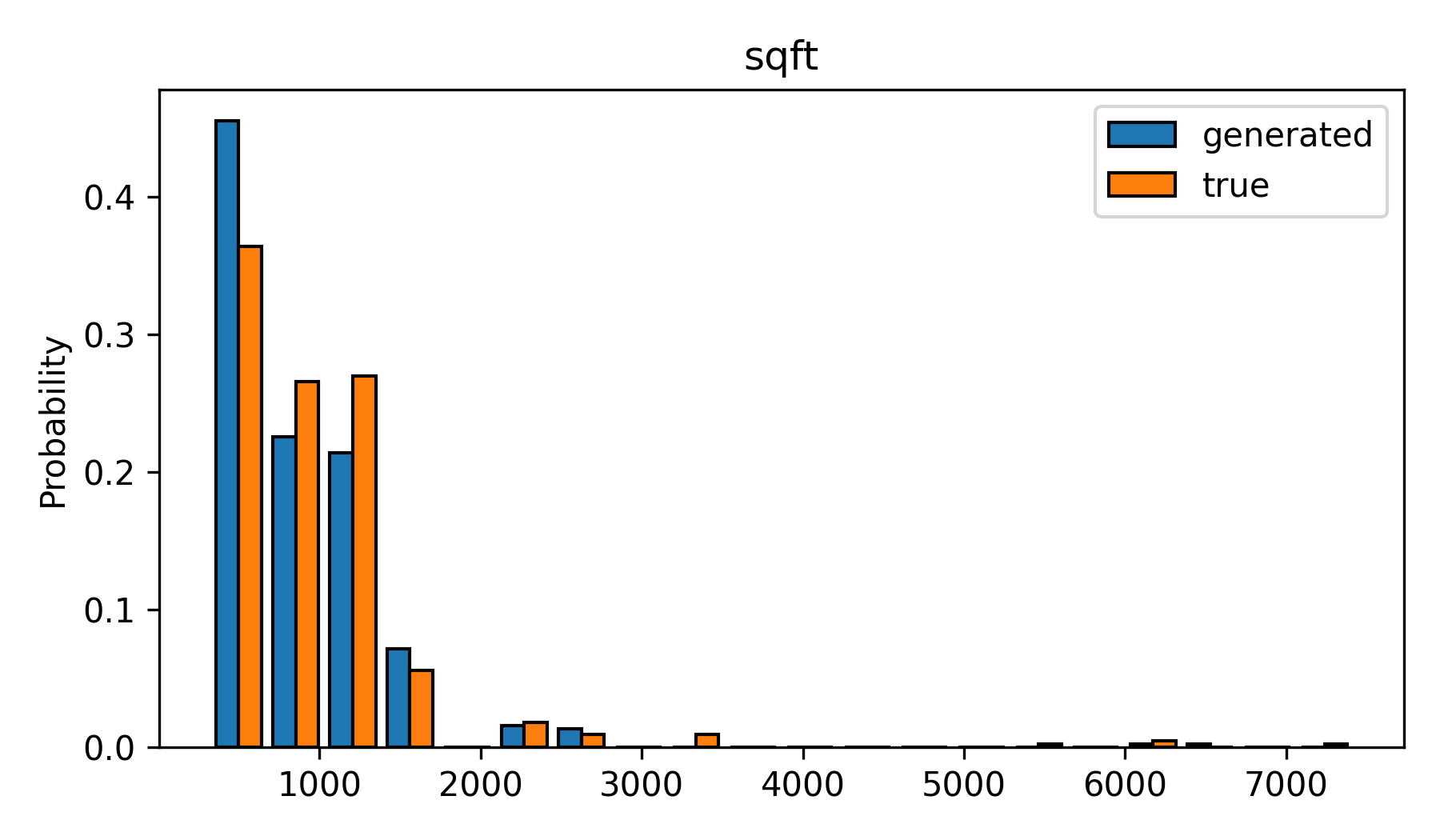}
        \caption{Building square footage (Wasserstein distance for this plot: 0.014). Key dependencies used here: \textit{PUMA = '25003301'}. }
    \end{subfigure}
    \hfill 
    \begin{subfigure}[b]{0.48\textwidth}
        \centering
        \includegraphics[width=\textwidth]{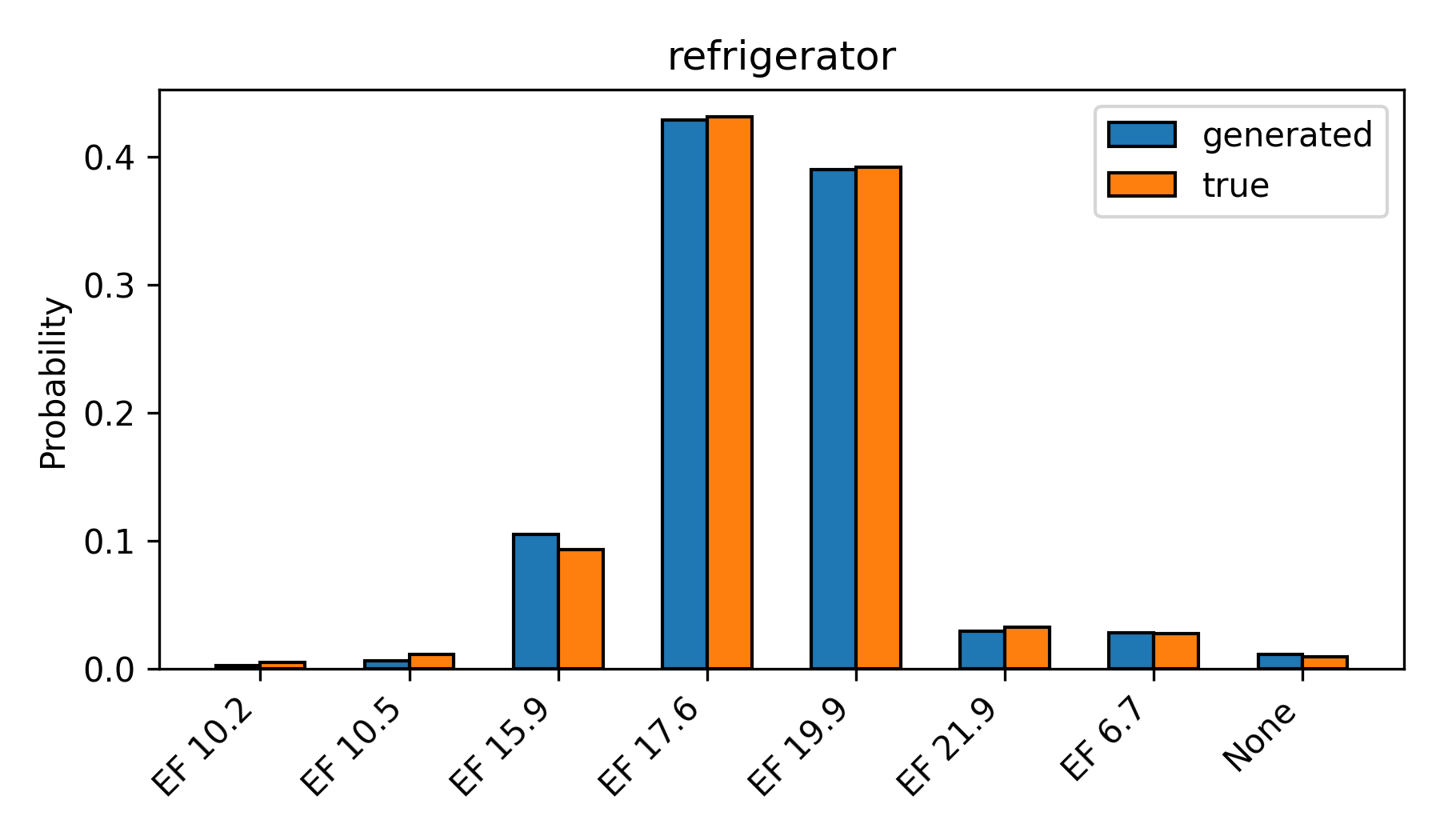}
        \caption{Refrigerator appliance (JS distance for this plot: 0.038). Key dependencies used here: \textit{building type = 'Single-Family Attached'} and \textit{vintage = '1940s'}.}
    \end{subfigure}

    \vspace{0.5cm} 
    \begin{subfigure}[b]{0.48\textwidth}
        \centering
        \includegraphics[width=\textwidth]{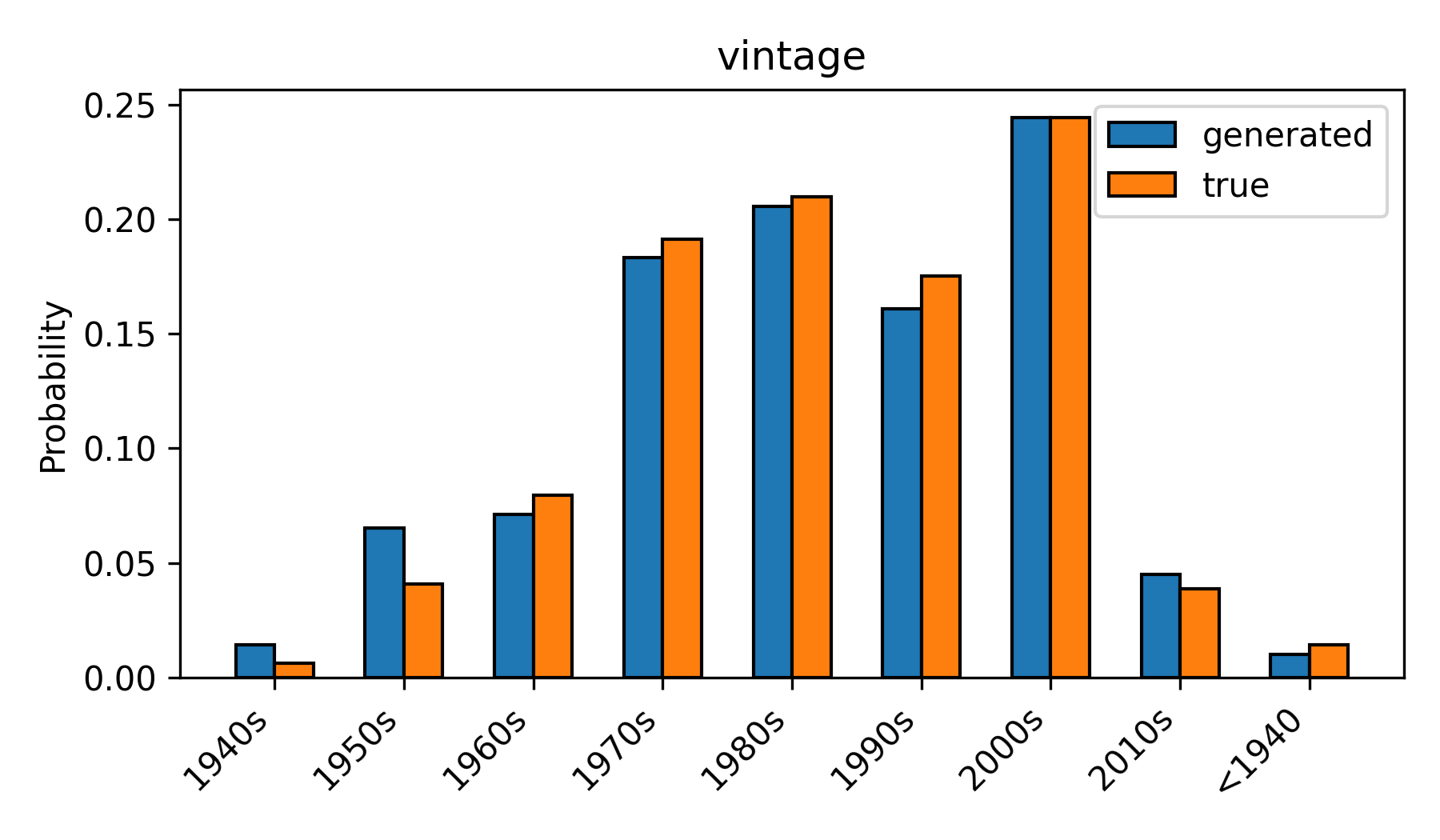}
        \caption{Building vintage (built year range) (JS distance for this plot: 0.064). Key dependencies used here: \textit{PUMA = '22002201'}.}
    \end{subfigure}
    \hfill 
    \begin{subfigure}[b]{0.48\textwidth}
        \centering
        \includegraphics[width=\textwidth]{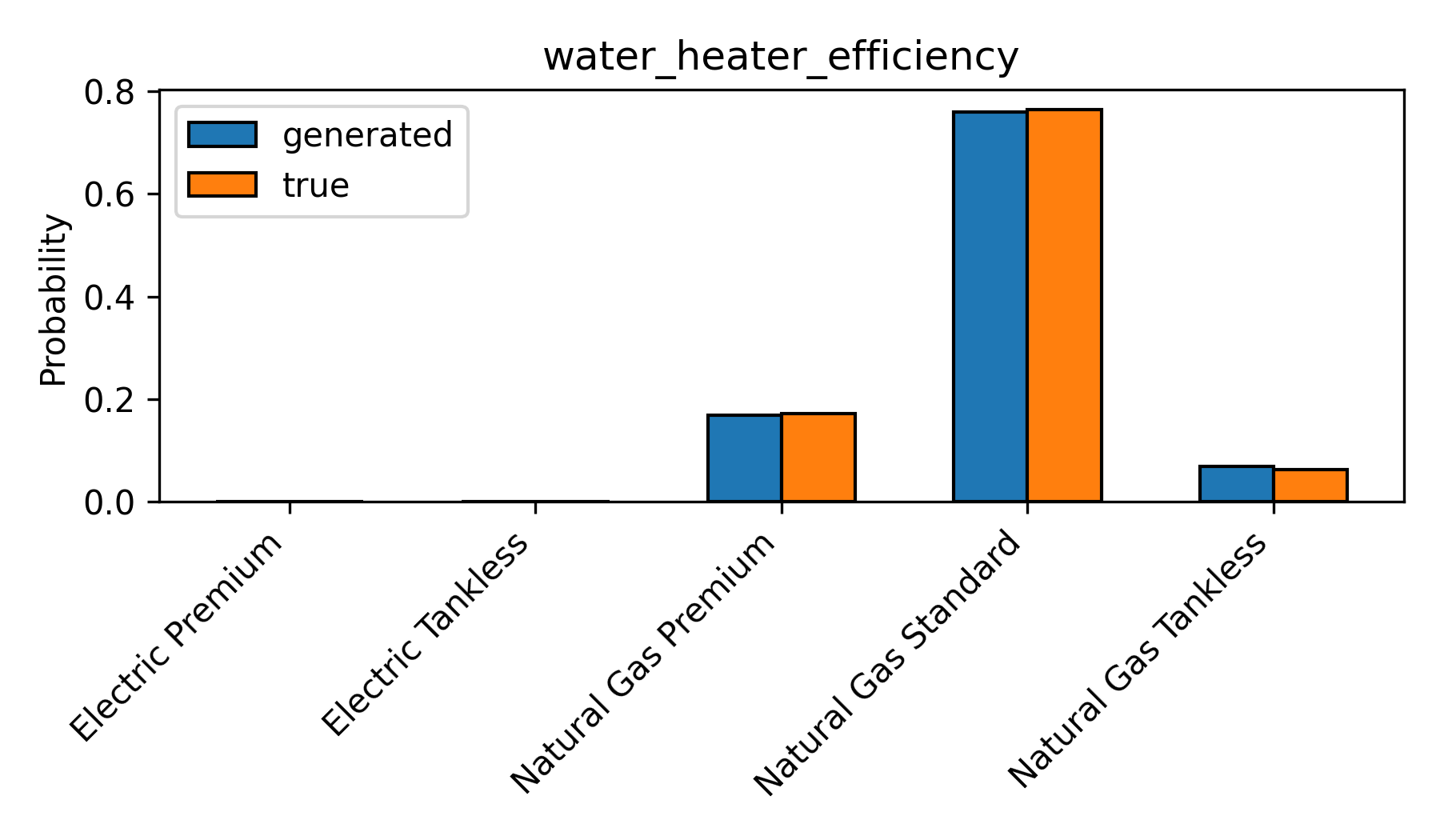}
        \caption{Water heater efficiency (JS distance for this plot: 0.033). Key dependencies used here: \textit{water heater fuel = 'Natural Gas'}.}
    \end{subfigure}

    \vspace{0.5cm} 
    \begin{subfigure}[b]{0.48\textwidth}
        \centering
        \includegraphics[width=\textwidth]{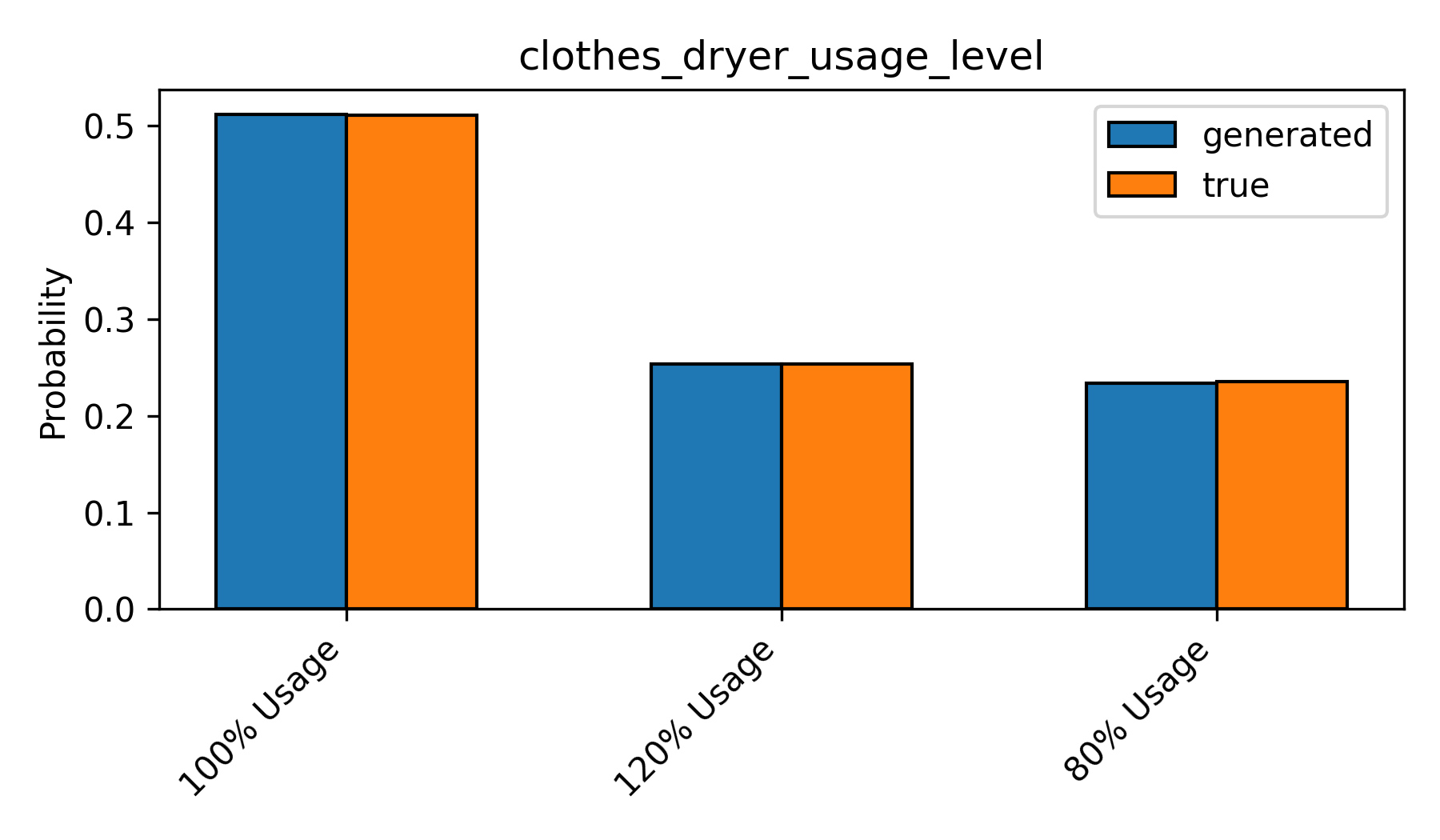}
        \caption{Clothes dryer usage level (JS distance for this plot: 0.001). No specified dependencies provided by ResStock.}
    \end{subfigure}
    \hfill 
    \begin{subfigure}[b]{0.48\textwidth}
        \centering
        \includegraphics[width=\textwidth]{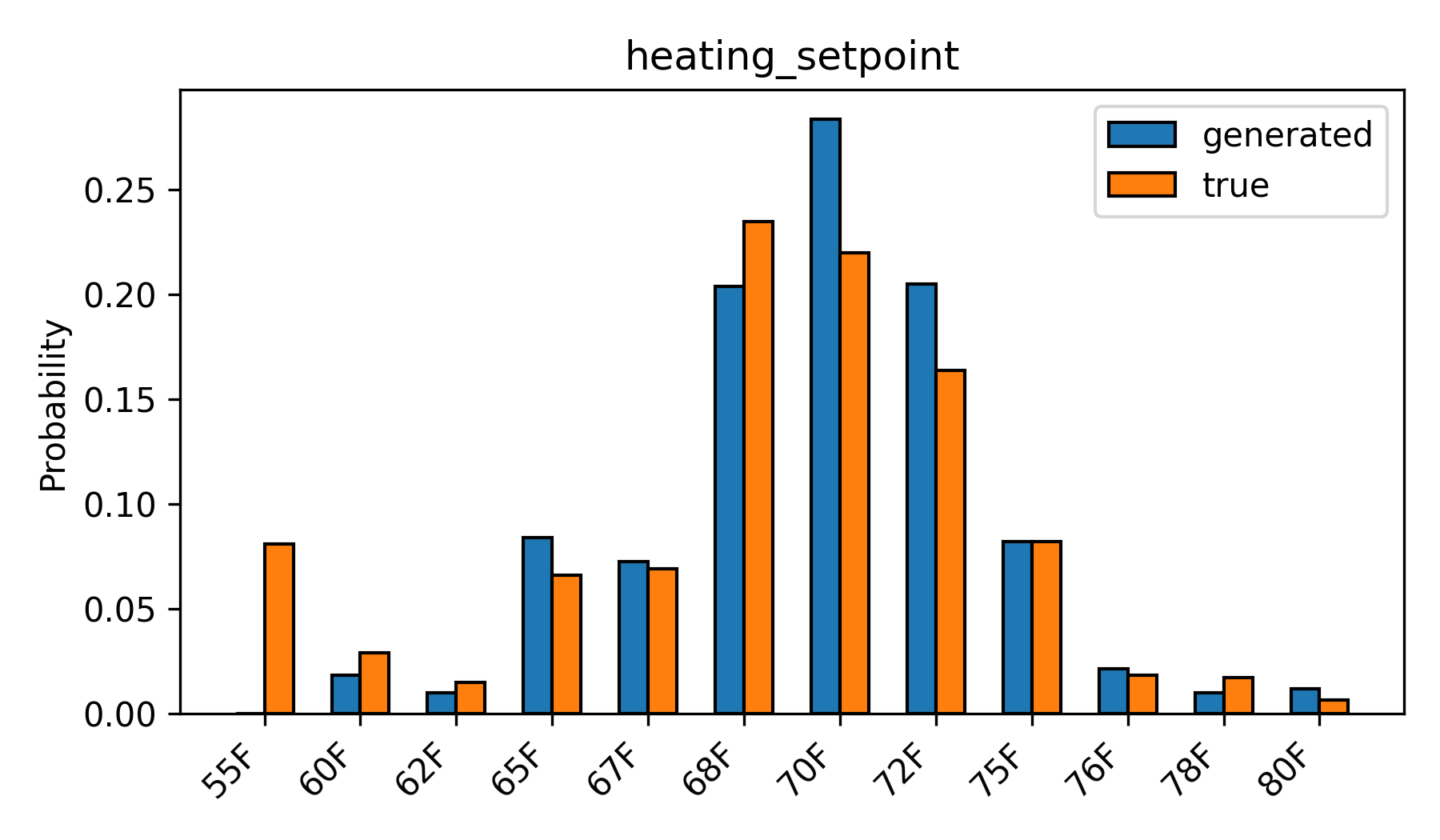}
        \caption{Heating setpoint (JS distance for this plot: 0.222). Key dependencies used here: \textit{HVAC has zonal electric (electric baseboard) heating = 'No'(Absent)} , \textit{building type = 'Single-Family Attached'} and \textit{HVAC heating type = 'Non-Ducted Heat Pump'}.}
    \end{subfigure}
    \caption{Comparison of the true vs generated univariate conditional distribution for a set of numerical and categorical features. We plot the distributions corresponding to the specific dependency combination (included in the sub-captions) that resulted in the minimum distance between the distributions. The results shown are obtained with the mixed imputation model.}
    \label{fig:univariate_comparison_plots}
\end{figure}
\\
\textbf{Result summary:}  The resulting true and generated distributions are visualized as histograms, and we show plots for both numerical and categorical features in Figure~\ref{fig:univariate_comparison_plots}. Quantitative distances are reported in Table~\ref{tab:univatiate_analysis}, showing an average over every unique dependency set for each variable. The Wasserstein distance is normalized by the range of the true samples. 
Table~\ref{tab:univatiate_analysis} includes results for the mixed-imputation model. We also compute 
results for one of the categorical-only imputation models, trained using a masking ratio of $0.05$ (which is equivalent to masking up to $2$ out of $31$ categorical features during training). The performance of the categorical-only imputation model remains consistent with that of the mixed imputation model, with an average JS-Distance of $0.197$ after averaging over all categorical variables. For comparison, we include a baseline that imputes each categorical variable by randomly choosing a category from all its possible categories. The JS distance for this baseline is averaged across all variables and reported in the last row of Table~\ref{tab:univatiate_analysis}, which shows the random baseline's poor performance as compared to our model. While deterministic predictive methods (such as gradient-boosted decision trees like XgBoost) are common tabular imputation baselines, they are unsuitable for our problem scope, as their fixed input-output mapping architectures would require training a computationally intractable number of distinct models to accommodate the arbitrary combinations of missing features present in a building characteristics dataset. Even the univariate and bivariate evaluations presented here on the ResStock dataset would require training hundreds of individual models, a complexity that scales into the millions as more features are simultaneously masked in real-world urban energy modeling setups (as seen in our case study). 

\begin{figure}[!htbp]
    \centering 
    \begin{subfigure}[b]{0.48\textwidth}
        \centering
        \includegraphics[width=\textwidth]{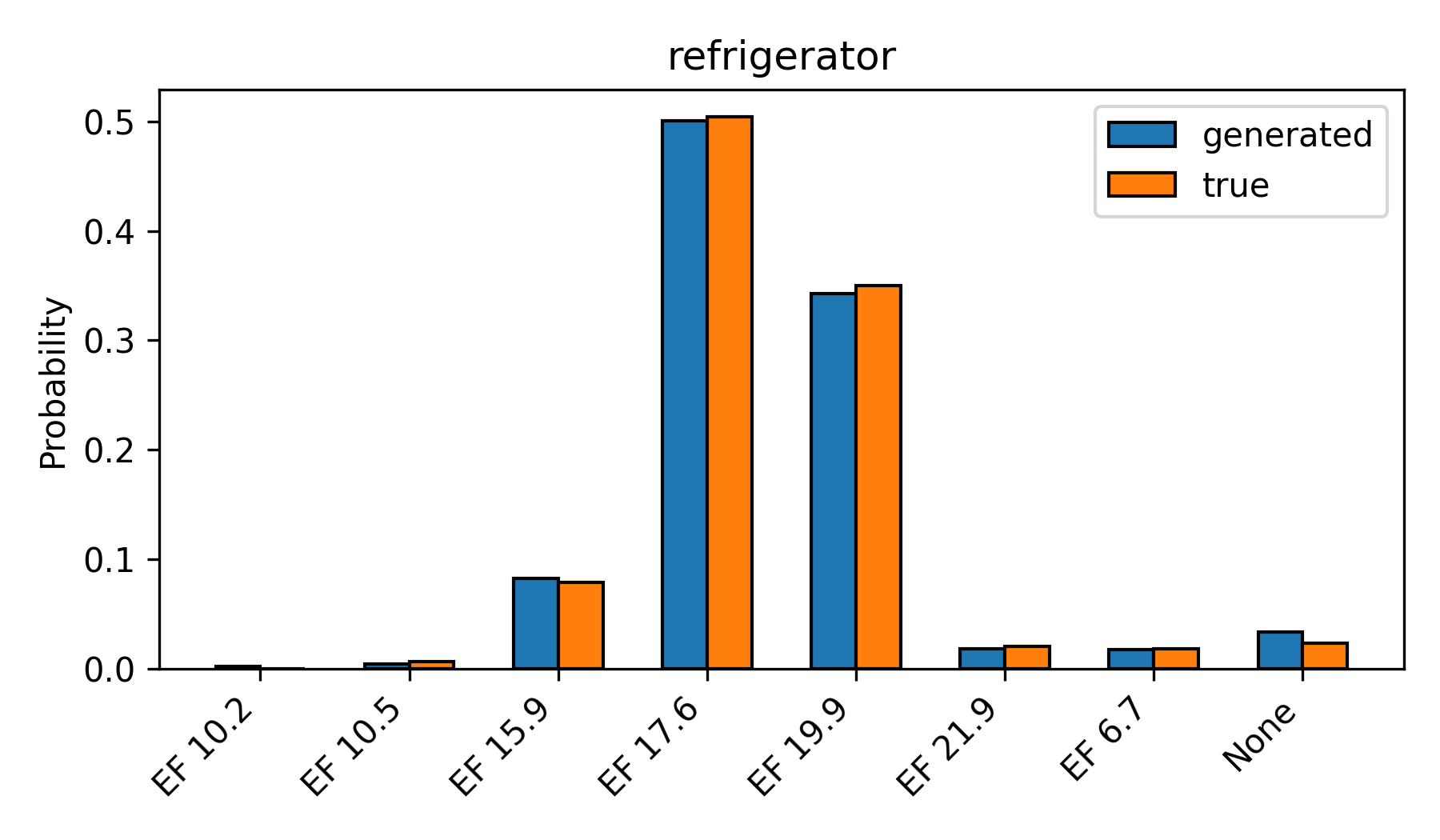}
        \caption{Refrigerator appliance (JS distance for this plot: 0.043). Key dependencies used here: \textit{building type = 'Multi-Family with 2 - 4 Units'} and \textit{vintage = '1990s'}.}
    \end{subfigure}
    \hfill 
    \begin{subfigure}[b]{0.48\textwidth}
        \centering
        \includegraphics[width=\textwidth]{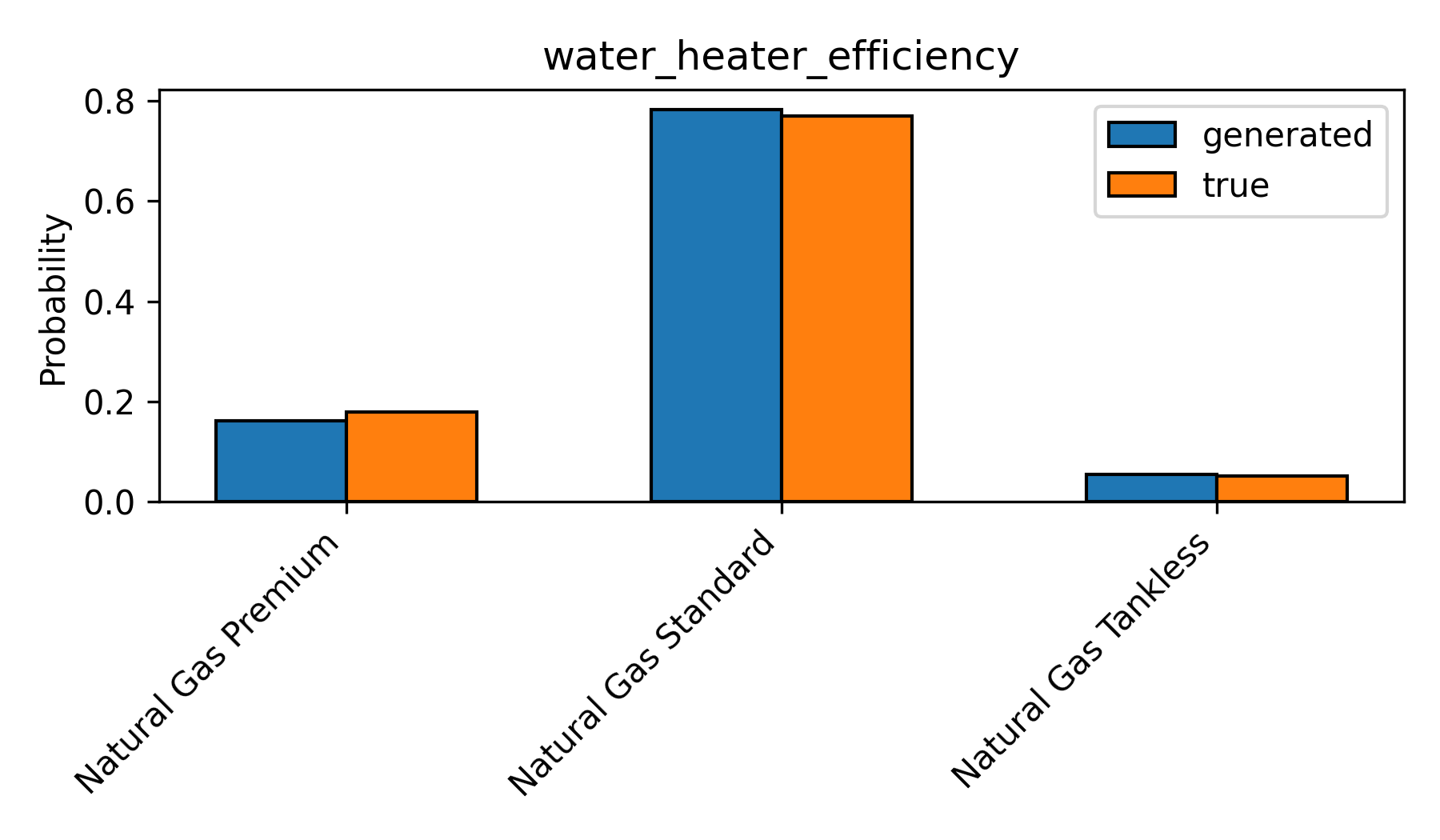}
        \caption{Water heater efficiency (JS distance for this plot: 0.021). Key dependencies used here: \textit{water heater fuel = 'Natural Gas'}.}
    \end{subfigure}

    \vspace{0.5cm} 
    \begin{subfigure}[b]{0.48\textwidth}
        \centering
        \includegraphics[width=\textwidth]{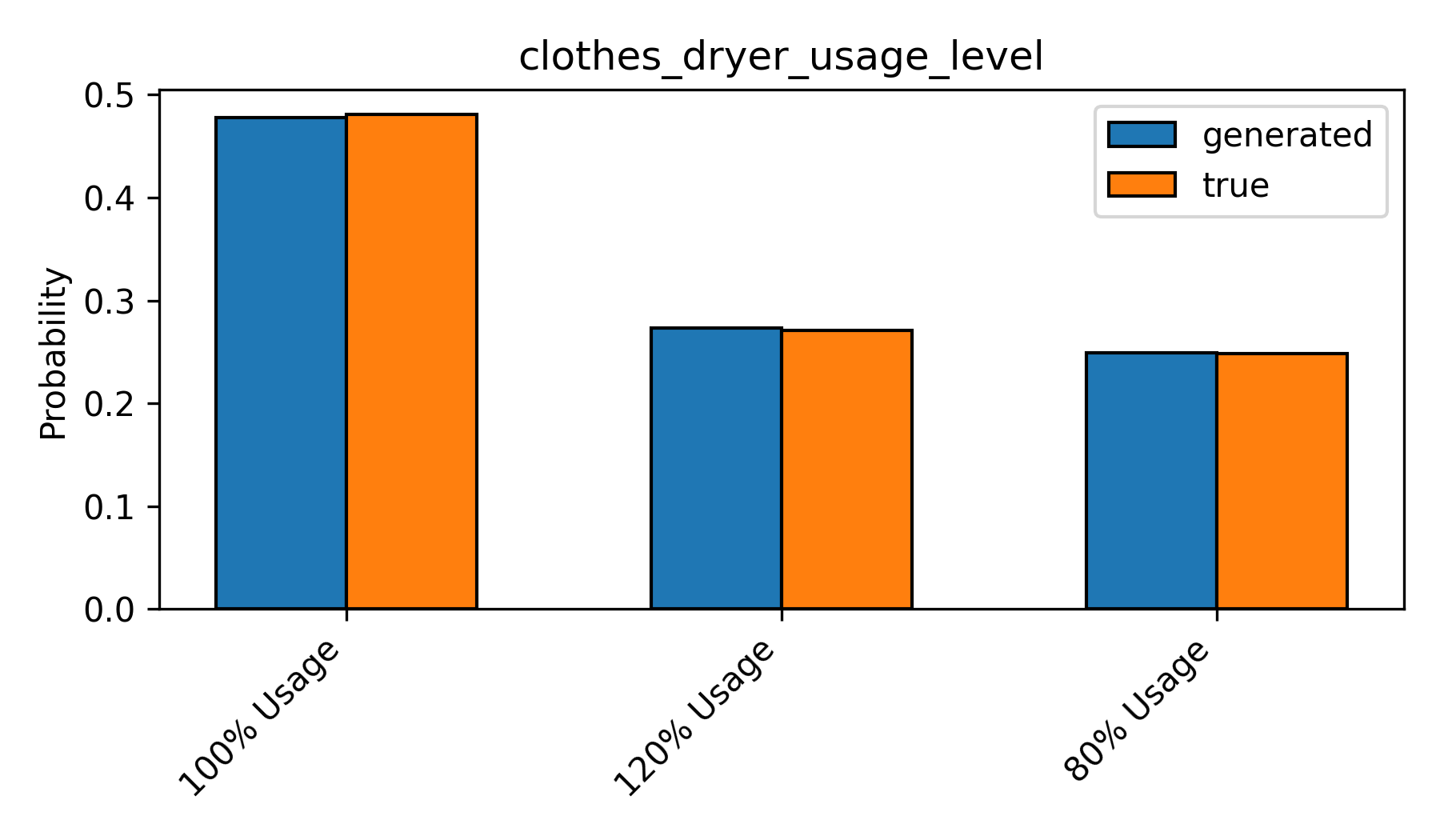}
        \caption{Clothes dryer usage level (JS distance for this plot: 0.003). No specified dependencies provided by ResStock.}
    \end{subfigure}
    \hfill 
    \begin{subfigure}[b]{0.48\textwidth}
        \centering
        \includegraphics[width=\textwidth]{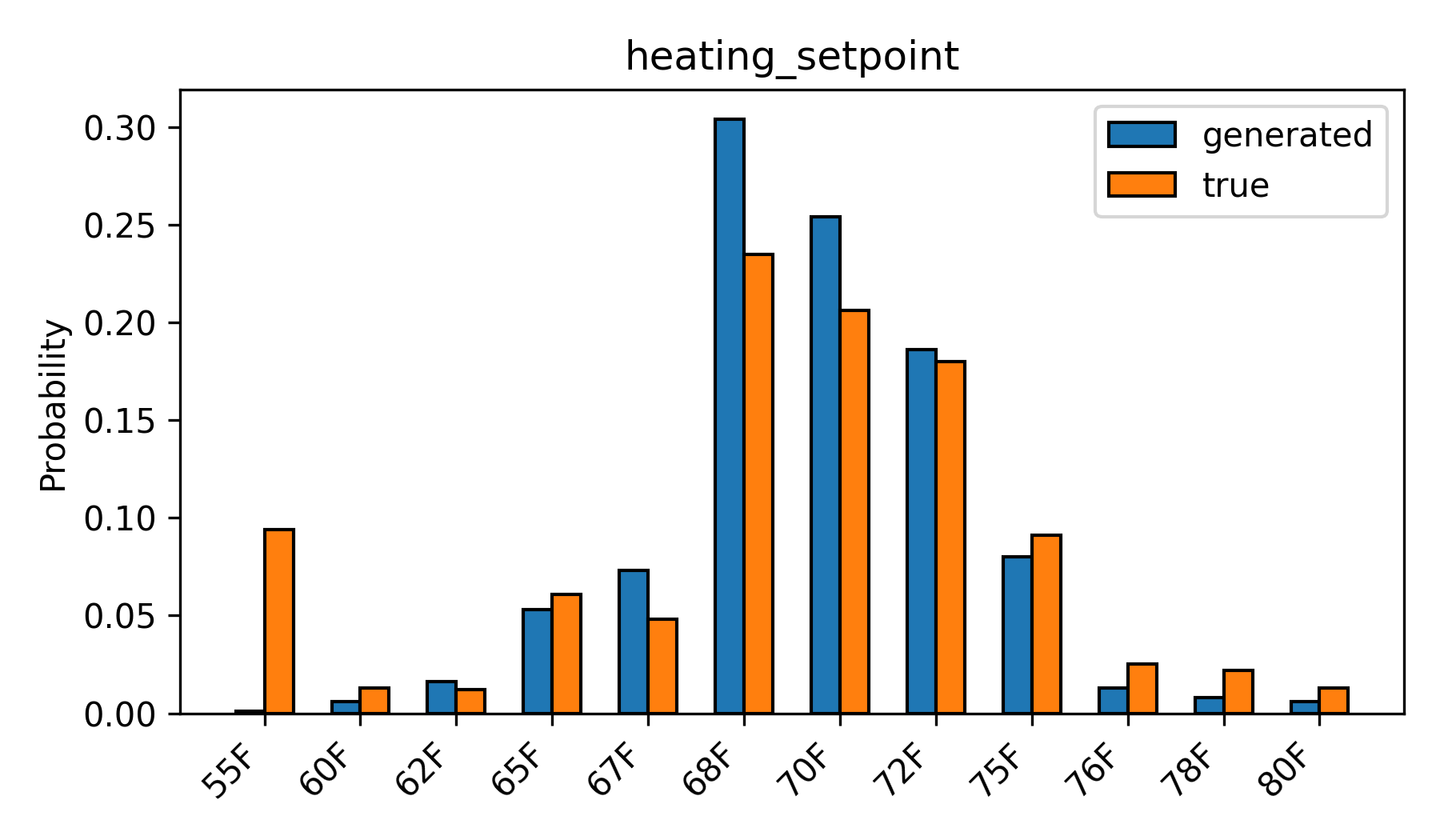}
        \caption{Heating setpoint (JS distance for this plot: 0.238). Key dependencies used here: \textit{HVAC has zonal electric (electric baseboard) heating = 'No'(Absent)}, \textit{building type = 'Single-Family Detached'} and \textit{HVAC heating type = 'Non-Ducted Heat Pump'}.}
    \end{subfigure}
    \caption{Comparison of the true vs generated univariate conditional distribution for a set of categorical features, evaluated on the OOD test dataset. We plot the distributions corresponding to the specific dependency combination (included in the sub-captions) that resulted in the minimum distance between the distributions. The results shown are obtained with the mixed imputation model.}
    \label{fig:univariate_comparison_plots_ood}
\end{figure}

We observe from Table~\ref{tab:univatiate_analysis} and Figure~\ref{fig:univariate_comparison_plots} that our model demonstrates strong conditional generation quality across many features, though the performance varies, with some variables performing better than others. According to Table~\ref{tab:univatiate_analysis}, some variables, such as heating/cooling Setpoint has offset, heating fuel, and building geometry type, are particularly
challenging for the model to learn. This may be due to complex underlying class structures or imbalanced classes for some of these variables~\cite{el5264131ai}. Figure~\ref{fig:univariate_comparison_plots} shows that the model's output is realistic and matches the true distribution closely, even with features having a higher number of categories, e.g., for refrigerator, vintage (built year range), and water heater efficiency; the generated distribution also captures the dominant modes well, like "Natural Gas Standard" for water heater efficiency. Notably, the model performs well even on features that lack strongly defined dependencies in the ResStock data, such as the clothes dryer usage level variable. The model is also effective in capturing the wide range of values present in the numerical distribution, e.g, for square footage. 
\\
\\
\textbf{Analysis on OOD:} Here, we validate our model's performance on the out-of-distribution (OOD) test dataset. Overall, we expect OOD conditional generation to be a challenging task, particularly for the variables that directly or indirectly depend on geospatial factors, as buildings in these PUMAs were not seen during training. The Figure~\ref{fig:univariate_comparison_plots_ood} shows generated vs true distribution plots on a subset of the variables visualized in Figure~\ref{fig:univariate_comparison_plots}. 
We see similar qualitative performance in Figure~\ref{fig:univariate_comparison_plots_ood} for all these variables. 
It is important to note that for variables such as vintage or building square footage, which are dependent on PUMA (as seen in the dependencies obtained from ResStock), a comparison to the true distribution is not possible since there are no matching training samples with the same PUMA. However, this challenge highlights the advantage of our approach - learning a single generative model on the entire U.S. residential building stock. This enables our model to still generalize from broader patterns and impute building characteristics for geographical locations unseen during training, unlike studies that train a separate 
neural network model for each district (\cite{el2024open,el5264131ai}).
\subsubsection{Bivariate analysis:} 
\begin{figure}[p]
 \caption{Comparison of the true vs generated bivariate conditional distributions. The central panel shows a heatmap of the error in the joint bivariate distribution, and the top and side panels compare the generated and true marginal distributions. We plot the distributions corresponding to the specific dependency combination that resulted in the minimum distance between the joint distribution. The results shown are obtained with the categorical-only imputation model trained using a masking ratio of $0.05$.}
 \vspace{-1ex}
    \centering 
    \begin{subfigure}[b]{0.65\textwidth}
        \centering
        \includegraphics[width=\textwidth]{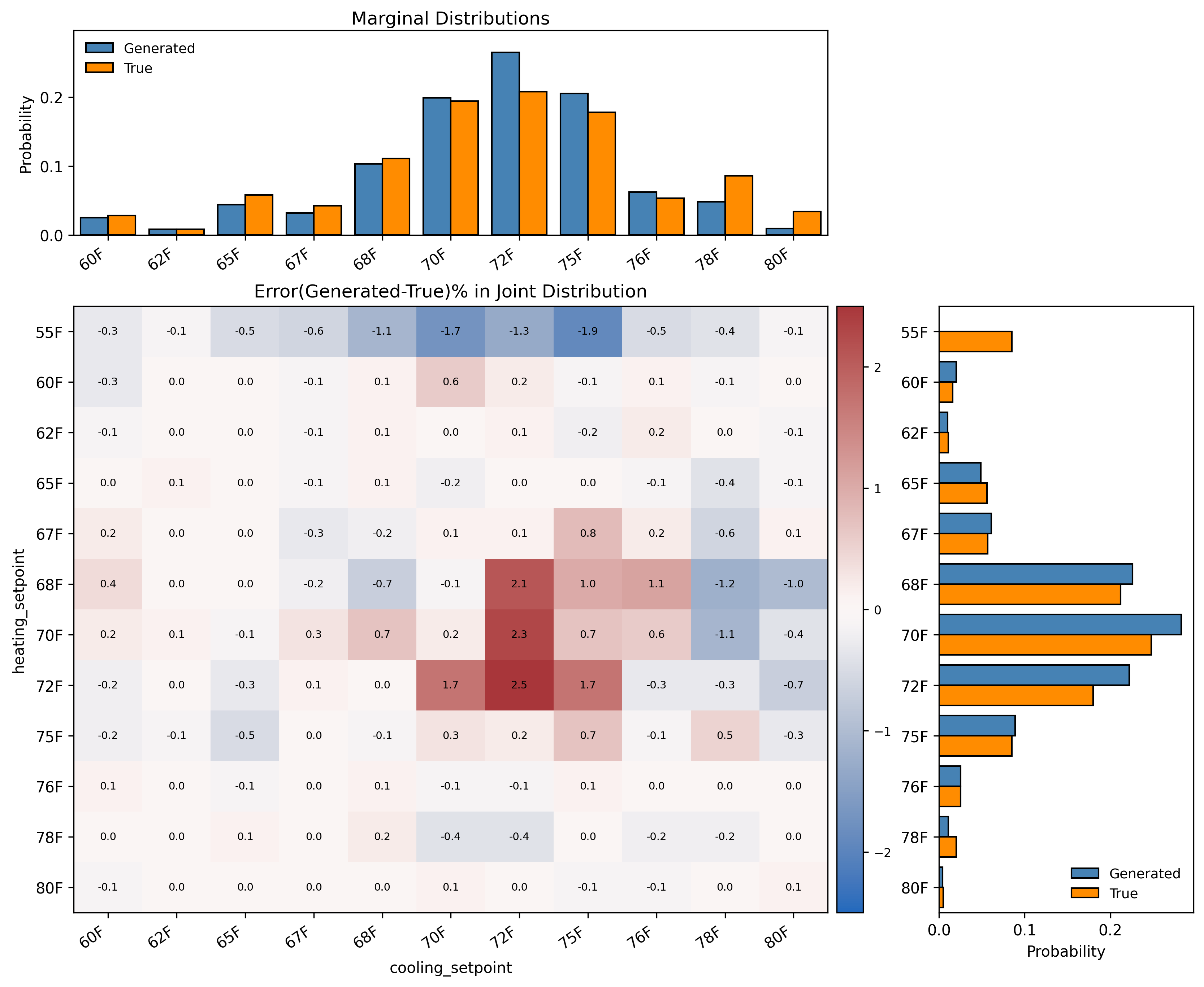}
        \caption{Heating setpoint and Cooling setpoint (joint JS distance for this plot: 0.305). Key dependencies used here: \textit{HVAC cooling type = 'Central AC'}, \textit{HVAC heating type = 'Non-Ducted Heating'}, \textit{HVAC has zonal electric (electric baseboard) heating  = 'No' (Absent)}, and \textit{building type = 'Single-Family Attached'}.}
    \end{subfigure}
    \begin{subfigure}[b]{0.65\textwidth}
        \centering
        \includegraphics[width=\textwidth]{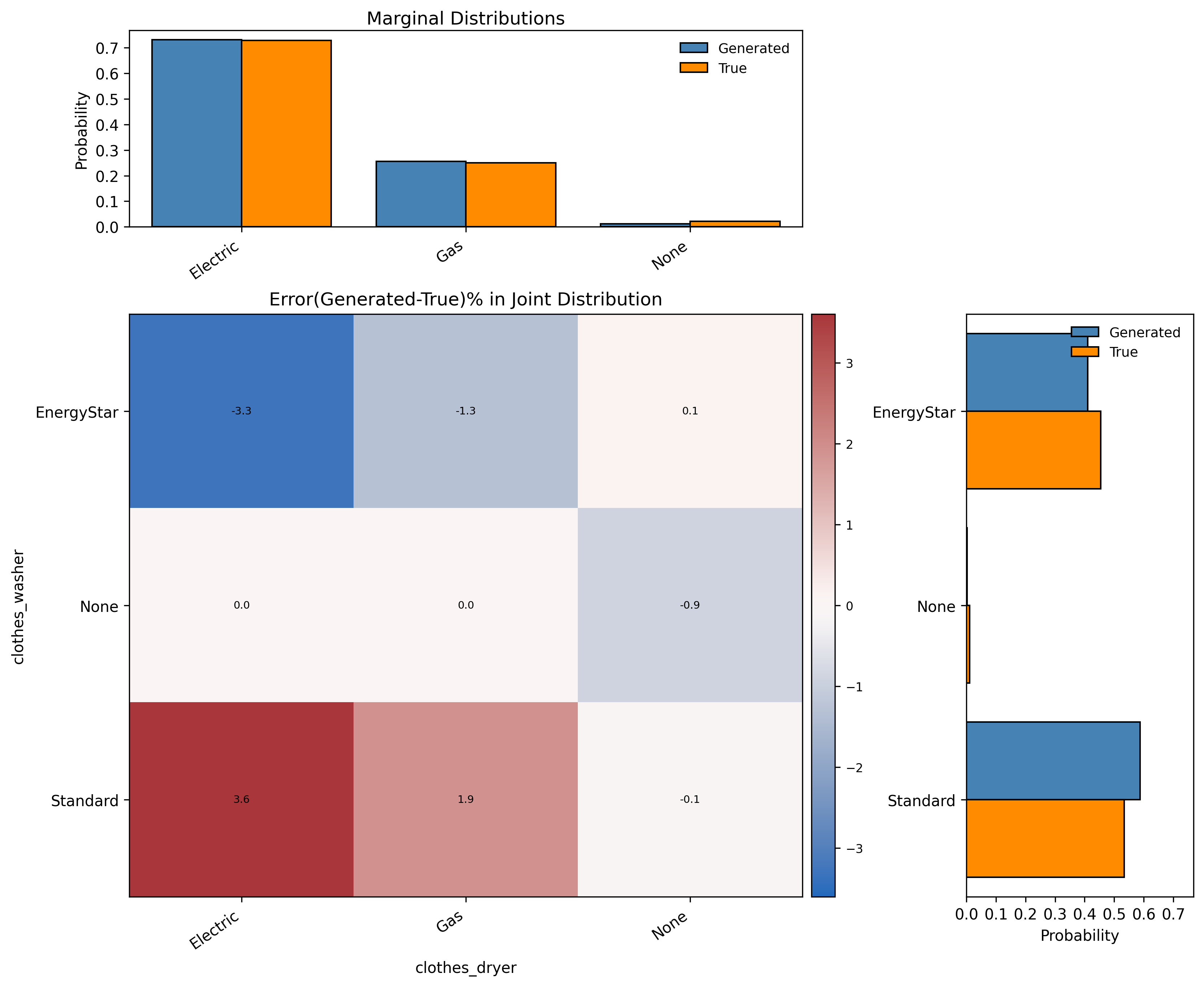}
        \caption{Clothes dryer and Clothes washer appliances (joint JS distance for this plot: 0.067). Key dependencies used here: \textit{heating fuel = 'Natural Gas'}, \textit{vintage  = '2010s'}, and \textit{building type = 'Single-Family Detached'}.}
    \end{subfigure}
   
    \label{fig:bivariate_analysis}
\end{figure}

Furthermore, we evaluate our model's ability to perform bivariate conditional generation, i.e, simultaneously imputing two variables given the rest of the variables, \( p(x_i^{\text{gen}},x_j^{\text{gen}} \mid x^{\text{cond}} )\). This is more challenging than generating realistic univariate distributions, as our conditional TabDDPM model also needs to capture interdependencies between the two variables. We begin by masking two features in the test dataset and follow a similar approach to obtain the joint true and generated distributions. We show this capability through Figure~\ref{fig:bivariate_analysis}, on two different pairs of features. The central panel shows the error in the joint bivariate distribution, while the panels at the top and to the side compare the generated and true marginal distributions for the two variables under study. The joint JS distance between the two distributions is also reported. We notice that our model performance for this complex task depends on the intricacy of the joint distribution. For example, for clothes washer and dryer (Figure~\ref{fig:bivariate_analysis}(b)), which are strongly related, the model shows smaller errors and JS distance, and better marginal distribution comparisons; however, with more complex and nuanced dependencies and a larger number of categories per feature, such as in heating and cooling setpoint (Figure~\ref{fig:bivariate_analysis}(a)), we see worse performance. While we don't extend this type of analysis for multivariate conditional generation (i.e, for more than two variables), we examine the imputation of up to 10 masked variables in the case study evaluation presented in~\ref{case_study}.

\subsection{Reconstruction evaluation}
\label{reconstruction_evaluation}
As an additional evaluation of the Conditional TabDDPM modeling framework, we compute the error between the generated synthetic outputs and the true feature values in our test dataset. While the analyses in~\ref{conditional_generation_evaluation} assess the realistic quality of the generated distribution, it is important to evaluate the model's ability to generate accurate point estimates.
Specifically, we systematically mask each imputable variable, $x_i$, in the test dataset, one at a time. We then generate imputed \(x_i^{\text{gen}}\) values using the Conditional TabDDPM model conditioned on the rest of the variables, which are observed. We measure the reconstruction score between the generated and ground truth value $x_i$ from test data, using root-mean-squared-error (RMSE) for numerical features and Accuracy score for the categorical features. The RMSE scores are normalized with the range of the ground truth column values. 

Table~\ref{tab:reconstruction_analysis} 
shows the reconstruction scores, averaged over all variables (of the same type). The Table includes results for two of our trained models: the mixed imputation model 
, and the 
categorical-only imputation model trained using a masking ratio of $0.05$. 
For a comprehensive analysis, we include baselines for comparison. We follow the steps from the univariate analysis; for every row in the test data, we find matching rows in the training set having the same set of dependency values. The baseline is then the mean of the target value from matching rows for numerical features, or the majority category (mode) for the categorical features. We also add another simpler baseline, only for categorical features, which is a randomly chosen category from all possible categories for the particular variable. The results from the table show the superior performance of our models, achieving lower RMSE and higher Accuracy over all the baselines. We also compute our models' reconstruction scores on the OOD test data. We observe a slightly higher RMSE of $0.092$ and a comparable Accuracy of $0.758$ with the mixed imputation model, and a comparable Accuracy of $0.761$ with the categorical-only imputation model.
\begin{table}[htbp]
\centering
\caption{\textbf{Reconstruction results} The table reports RMSE for numerical features and Accuracy for categorical features, averaged over all features and over three random seeds. The lower the RMSE, the better, and the higher the Accuracy, the better. We show the performances of two models, the mixed and the categorical-only imputation model. These models' reconstruction scores are compared to the baselines.}
\label{tab:reconstruction_analysis}
\begin{adjustbox}{max width=\columnwidth}
\begin{tabular}{l c c | c c c}
\hline 
& \multicolumn{2}{c}{\bfseries Model Performance} & \multicolumn{3}{c}{\bfseries Baseline Performance} \\
\cline{2-3} \cline{4-6} 
{\bfseries Model} & {\bfseries RMSE ↓} & {\bfseries Accuracy ↑} & {\bfseries RMSE ↓} & {\bfseries Random Accuracy ↑} & {\bfseries Mode Accuracy ↑} \\
& {\bfseries (Num)} & {\bfseries (Cat)} & {\bfseries (Num)} & {\bfseries (Cat)} & {\bfseries (Cat)} \\
\hline 
Mixed imputation        & 0.076 & 0.759 & 0.099 & 0.263 & 0.644 \\
\rule{0pt}{2.5ex} %
Cat-only imputation    & ---  & 0.762 & --- & 0.263 & 0.644  \\
\hline
\end{tabular}
\end{adjustbox}
\end{table}

\subsection{Case Study}
\label{case_study}
\begin{figure}[!htbp]
    \centering
    \includegraphics[width=0.4\textwidth]{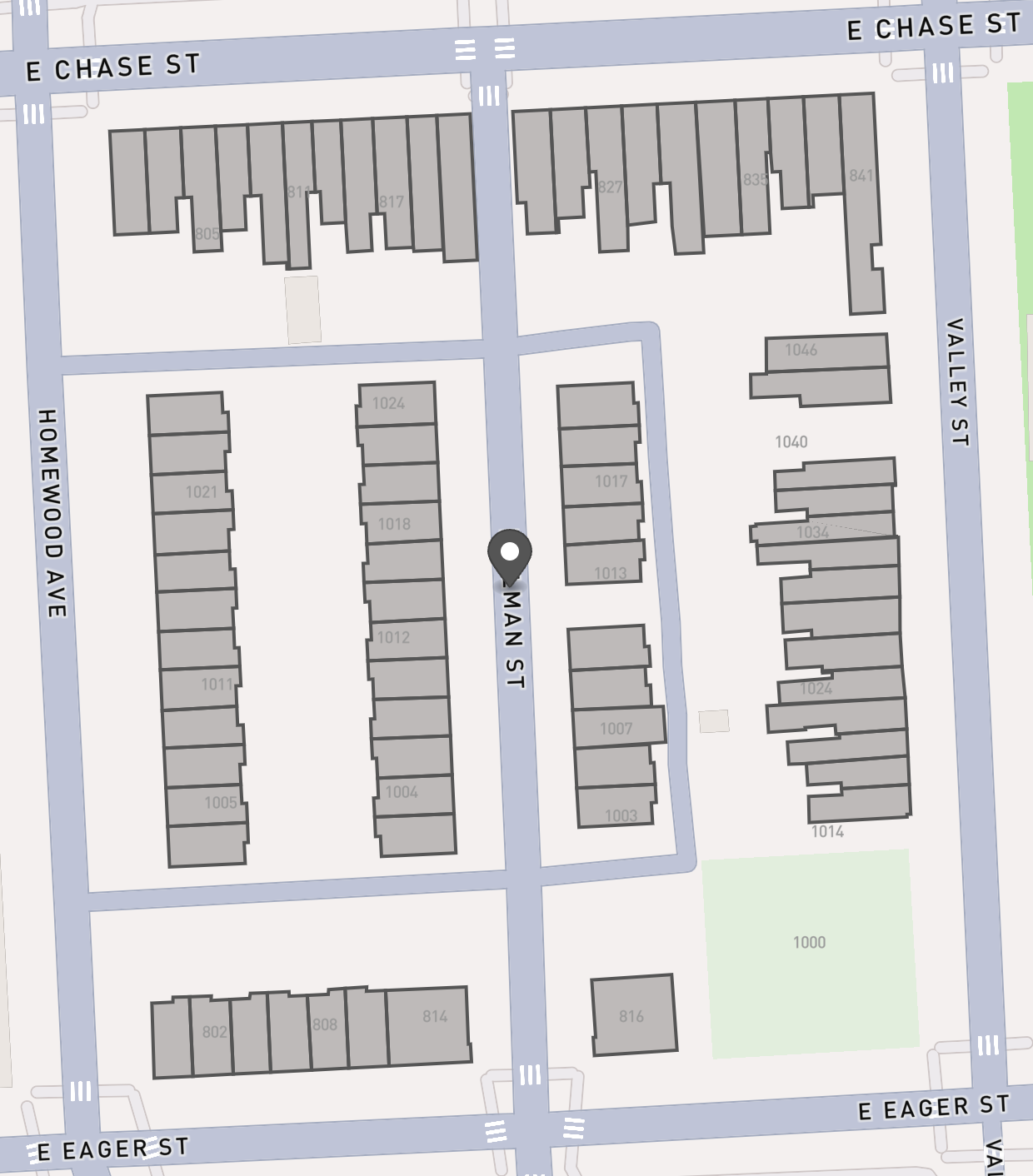}
    \caption{
    A residential community in Baltimore is used in our case study (based on~\cite{el5264131ai}). There are $77$ single-family attached buildings. Building characteristics such as building type, attic type, foundation type, building area (square footage), built year, number of stories, and heating fuel are obtained for these buildings from sources like OpenStreetMap and Zillow/Redfin. 
    }
    \label{fig:case_study}
\end{figure}

We demonstrate the application of our framework in a case study on a residential neighborhood in Baltimore, Maryland (see Figure~\ref{fig:case_study}). Our study integrates our conditional generative modeling pipeline with the URBANopt~\cite{el2020urbanopt,polly2016zero} energy model following the setup in Kontar et al.~\cite{el5264131ai}. Evaluating our model's performance at the community level is the most relevant metric for our intended application, as it allows us to capture energy use patterns at this scale and supports district-scale planning.  Specifically, given a set of observed characteristics for the $77$ buildings in this neighborhood, we employ our Conditional TabDDPM model to impute all the unknown building characteristics. This synthetically completed data is then passed as an input to URBANopt, a physics-based energy model, which simulates load profiles in terms of energy or electricity consumption per building.

\begin{figure}[!htbp]
  \centering
  \setlength{\tabcolsep}{0pt}
  \begin{tabular}{@{} m{0.74\textwidth} @{\hspace{0.5em}} m{0.22\textwidth} @{}}
    \includegraphics[width=\linewidth,trim=0 0 0 12pt,clip]{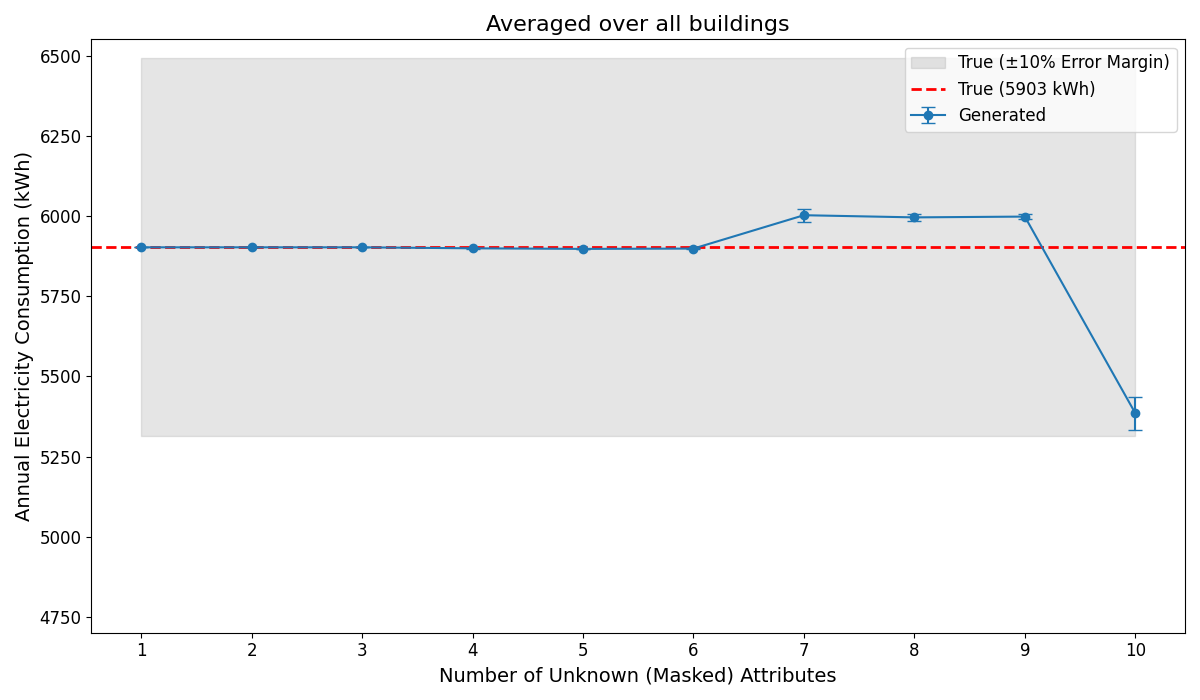} &
    \centering\includegraphics[width=0.8\linewidth]{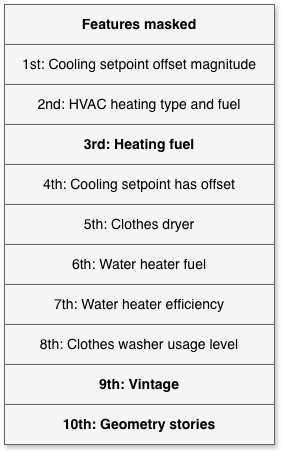}
  \end{tabular}

  \caption{
    Annual electricity consumption (in kWh) simulated by URBANopt, and averaged across all buildings. The plot (on the left) compares the simulation results obtained using the reference building characteristics ("True")  against those generated by the Conditional TabDDPM model ("Generated"). The grey-shaded region represents the $\pm 10\%$ error margin around the "True" value (red dashed line). Our model imputes varying levels (up to 10) of unknown or masked characteristics as shown on the x-axis, with the features being masked cumulatively in the order listed in the table to the right. We sample $5$ times from our model to generate multiple scenarios of plausible building characteristics based on the same conditions. The "Generated" points are the mean of the simulation outputs, with the error bars showing the standard deviation across these scenarios. These results are for the categorical-only imputation model, trained using a masking ratio of $0.4$. 
    }
    \label{fig:case_study_results}
\end{figure}

\begin{figure}[!htbp]
  \centering
  \setlength{\tabcolsep}{0pt}
  \begin{tabular}{@{} m{0.74\textwidth} @{\hspace{0.5em}} m{0.22\textwidth} @{}}
    \includegraphics[width=\linewidth,trim=0 0 0 12pt,clip]{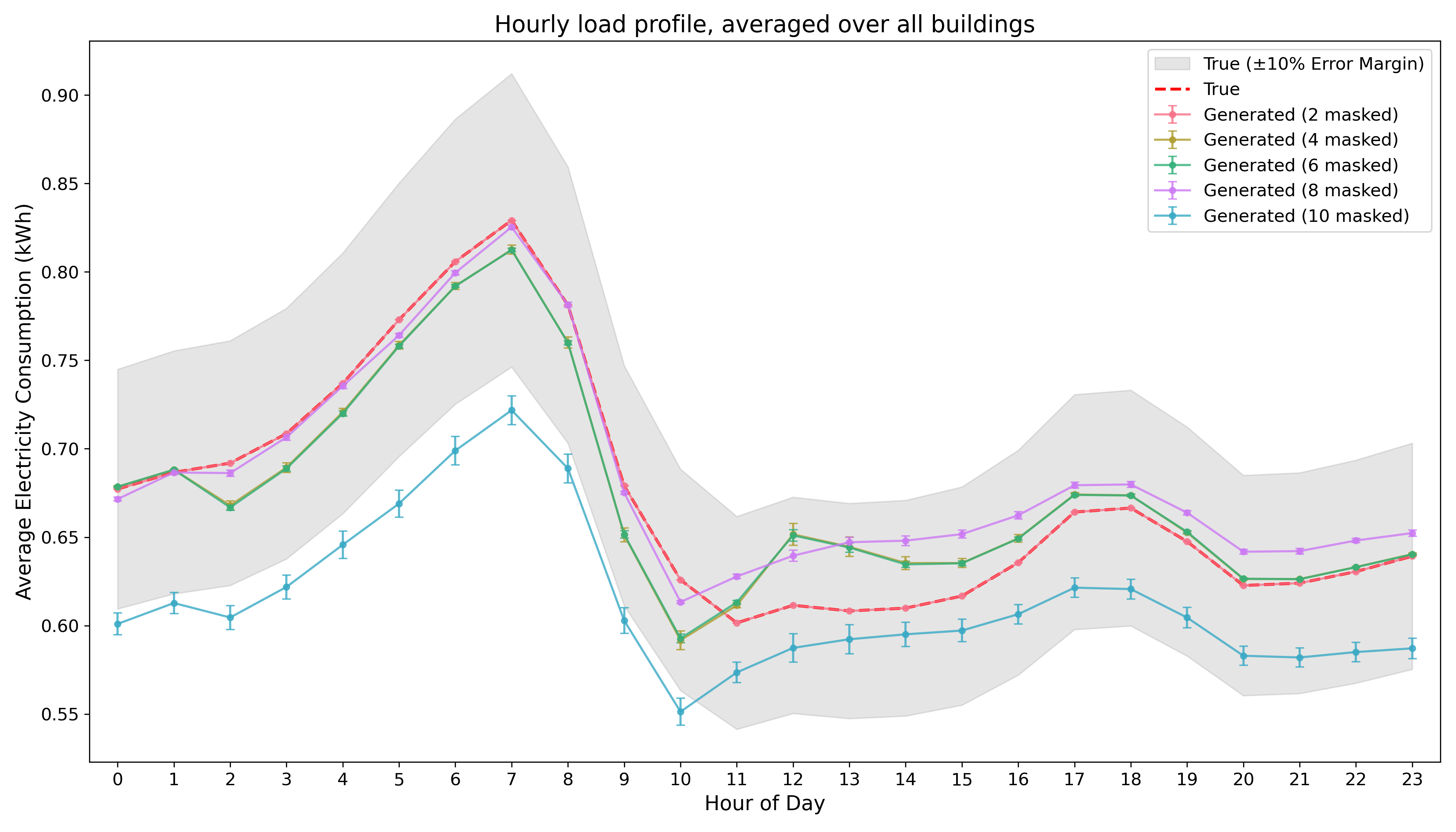} &
    \centering\includegraphics[width=0.8\linewidth]{masked_table_updated_with_bold.png}
  \end{tabular}
  \caption{
    Annually averaged 24-hour electricity consumption profile (in kWh), simulated by URBANopt, and averaged across all buildings. The plot (on the left) compares the timeseries simulation outputs obtained using the reference building characteristics ("True")  against those generated by the Conditional TabDDPM model ("Generated"). The grey-shaded region represents the $\pm 10\%$ error margin around the "True" profile (red dashed line). We plot generated profiles corresponding to scenarios with 2, 4, 6, 8, and 10 masked attributes. The table on the right lists the features that are masked and the order in which they are progressively masked.
    }
    \label{fig:case_study_time_series_results}
\end{figure}

We utilize building characteristics for the buildings in this neighborhood as curated in El Kontar et al.'s work~\cite{el5264131ai}, and we refer to this as the reference dataset. These include building attributes collected from open-source databases and characteristics produced by their neural network-based predictive model (discussed in~\ref{related_work_ml_for_buildings_application}). We emphasize that this reference dataset of building characteristics does not contain any actual metered data, as that would be extremely challenging to obtain. It is a hybrid of critical characteristics such as building type, attic type, foundation type, building square footage, vintage, geometric stories, and heating fuel  collected from real-world sources including OpenStreetMap and Zillow/Redfin, and the remaining characteristics predicted from the neural network model~\cite{el5264131ai}. From this set of characteristics, we select the $35$ building features in the Conditional TabDDPM model setup. We keep all the numerical features as observed and mask varying levels of categorical features. 
We start by masking one categorical feature, then two features, and then progressively increase the number of masked features to $10$, to assess our model's ability in imputing increasing levels and arbitrary combinations of unknown characteristics conditioned on the rest of the observed attributes. Crucially, we specifically included the masking of three real-world derived variables: vintage, geometric stories, and heating fuel (bolded in Figures~\ref{fig:case_study_results} and~\ref{fig:case_study_time_series_results}), to ensure we evaluate the model's ability in recovering such characteristics. The observed attributes used for conditioning notably always included the real-world derived building square footage,  as numerical features were kept fully observed in this case study.
After imputing the corresponding number of missing (masked) variables with our model, we evaluate the imputed features' quality by comparing URBANopt's resulting load profile using this generated dataset against the true load profile, which is simulated using the reference dataset from~\cite{el5264131ai}. 

Figure~\ref{fig:case_study_results} compares the annual total electricity consumption from the URBANopt output for both the true and generated setup, where the values are averaged across all buildings in the community. We observe that the conditional generation performance decreases slightly as the number of masked (unknown) attributes increases, and this drop becomes more pronounced when the model is tasked with imputing $10$ variables simultaneously, particularly since the final masked features: vintage and number of building stories, are critical variables in determining energy consumption. Given the probabilistic nature of the conditional generation, we sample $5$ times to generate multiple scenarios of plausible building characteristics based on the same set of observed conditions. The error bars in Figure~\ref{fig:case_study_results} represent the standard deviation in simulation output across these scenarios, and show that this variance increases slightly with the number of masked attributes. We include a $10\%$ error margin around the true line in the figure to visualize when the annual electricity consumption with generated data incurs an error greater than $\pm 10\%$. Notably, this doesn't occur, even for the highest number of masked attributes. 

In addition to comparing the annual electricity consumption, we extend our analysis to examining an hourly load profile or an hourly timeseries of electricity consumption as seen in Figure~\ref{fig:case_study_time_series_results}. 
The figure shows an annually averaged 24-hour profile of electricity consumption, comparing the URBANopt outputs for both the true and the Conditional TabDDPM generated setup, where the values are averaged over all buildings in the community. We plot generated profiles corresponding to scenarios with 2, 4, 6, 8, and 10 masked attributes. The simulated average electricity consumption using the generated data remains close in shape and magnitude (within the $\pm 10\%$ error margin) to the reference profile, at all hours of the day, except for the case with the highest number of masked attributes. This error, seen when imputing 10 unknown features, is most pronounced at the morning peak hour (7th hour), when we note a noticeable underestimation in the simulated consumption using our conditional generation model's output.

Overall, the most important observation is that the simulated hourly and annual electricity consumption with the generated characteristics stays close to the reference baseline, demonstrating the effectiveness of our conditional generation model even when almost a third of the total features are masked. This shows our model's potential in generating reliable and complete datasets, making it a valuable tool in end-to-end building energy modeling workflows.

\section{Discussion}
\begin{figure}[!htbp]
    \centering 
    \begin{subfigure}[b]{0.48\textwidth}
        \centering
        \includegraphics[width=\textwidth]{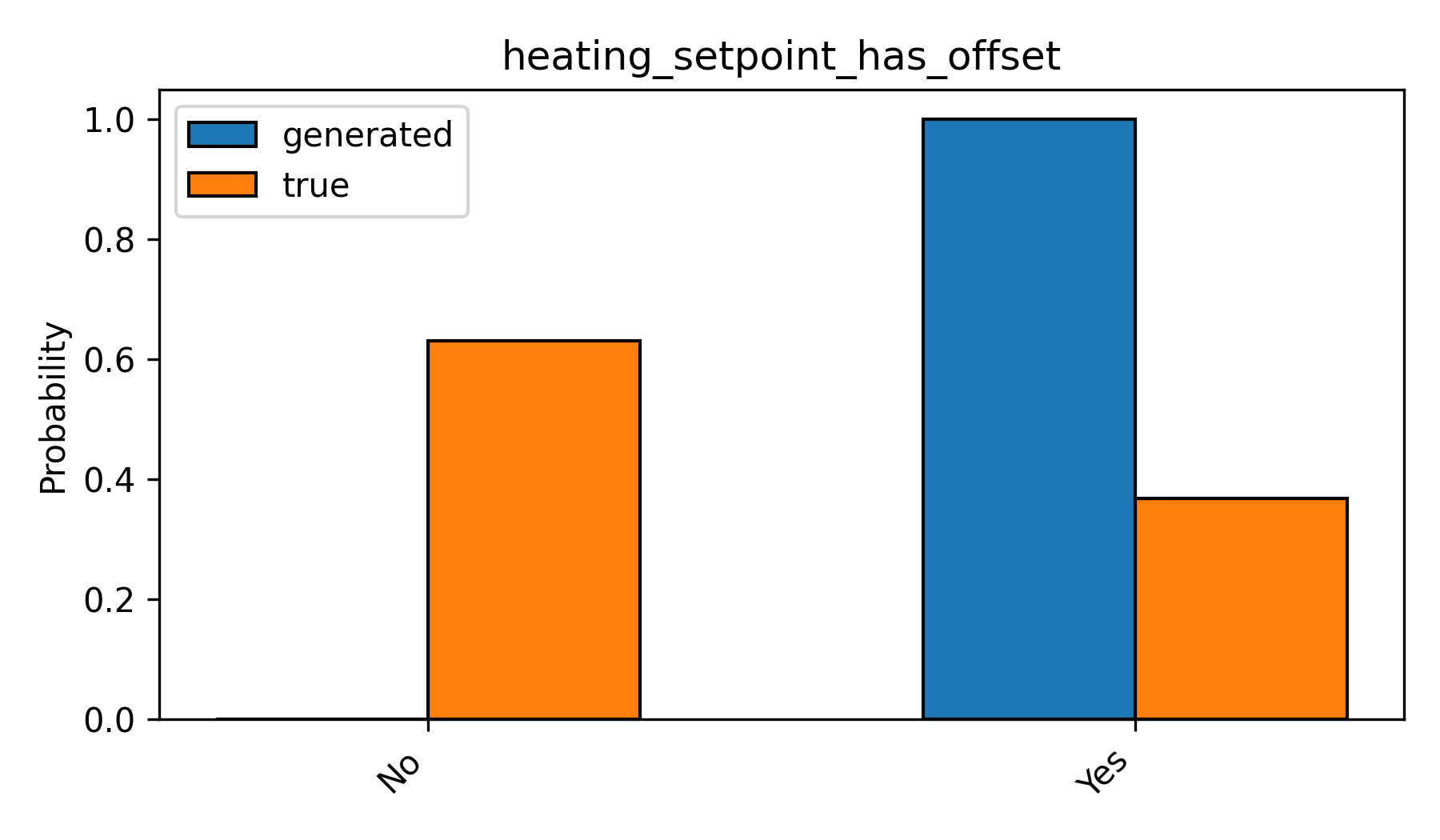}
        \caption{Heating setpoint has offset (JS distance for this plot: 0.651). Key dependencies used here: \textit{HVAC has zonal electric (electric baseboard) heating = 'No'(Absent)}, \textit{building type = 'Multi-Family with 2 - 4 Units’}.}
    \end{subfigure}
    \hfill 
    \begin{subfigure}[b]{0.48\textwidth}
        \centering
        \includegraphics[width=\textwidth]{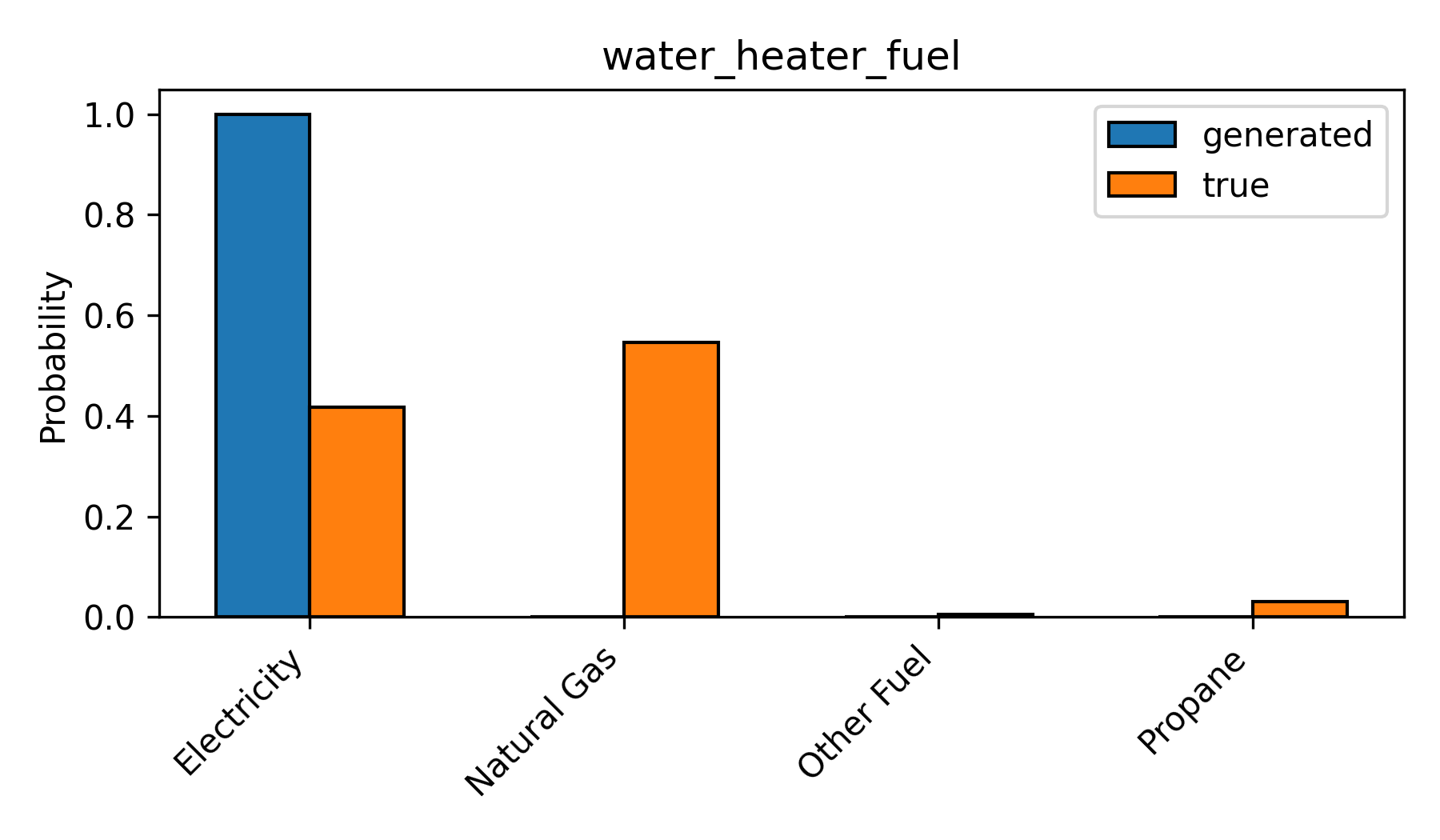}
        \caption{Water heater fuel (JS distance for this plot: 0.615). Key dependencies used here: \textit{heating fuel = 'None'}, \textit{building type = 'Single-Family Detached'}.}
    \end{subfigure}
    \caption{Comparison of the true vs generated univariate conditional distribution for two features, showing the model collapsing to a single generated category for a given set of observed conditions. The results shown are obtained with the mixed imputation model.}
    \label{fig:mode_collapse_plots}
\end{figure}

The results presented in the paper successfully validate the use of a conditional diffusion framework in generating realistic building attributes and creating complete characteristics in building-level datasets.
Our findings in Section~\ref{evaluation} demonstrate the model's ability to generate imputed features
whose distributions closely match the true ResStock distribution for many building variables. However, a feature-level breakdown of the JS distance between the two distributions in Table~\ref{tab:univatiate_analysis} shows that a few variables are more challenging to learn than others. It is possible that some of these features, like building types and the number of occupants, which are critical conditioning variables for generating other variables~\cite{el5264131ai}, are not strongly determined by other observed features in return. Moreover, detailed analysis of some variables, including a few with the worst performance, such as heating/cooling setpoint has offset, heating fuel, and water heater fuel,
reveals that our model learns an almost deterministic conditional mapping in these specific cases. The model consistently collapses to a single generated outcome when it is sampled conditioned on a given set of observed values, as seen in  Figure~\ref{fig:mode_collapse_plots}. This behavior, is characterized as conditional distribution mode collapse. This means that, while the generated values remain highly sensitive to the input variables, and vary as those conditions change, for a particular input the model produces a Dirac delta probability distribution (all mass at a single output class). While these variables perform great at single-point prediction, scoring high in reconstruction accuracy, the probabilistic nature of the outputs is not captured well in these cases. Although we observe this behavior to be limited to only a few variables, it is possible that an improved diffusion model workflow could address this in future work.

Overall, our evaluation focused on comparing the generated and true probability distributions in the univariate and bivariate conditional generation setting. While we don't extend this analysis to higher-dimensional multivariate cases, we demonstrate our model's multivariate conditional generation capabilities by simultaneously imputing up to $10$ building attributes, i.e, nearly a third of the total variables, in a real-world case study evaluation. Furthermore, our validation experiments with the OOD dataset show the model's strong performance in unseen geographical regions, which is an important advantage of our global generative modeling framework. Finally, we showcase an end-to-end application where our model generates a completed building-level dataset needed as an input to the URBANopt energy model. Our case study on a small residential community analyzes the simulated energy load profiles for these buildings, which closely match a curated (reference) baseline. 
This study underscores our model's potential in generating physically plausible and reliable datasets for practical use cases, showing its utility in enhancing building
energy modeling workflows. However, it is important to note that the generated outputs represent statistically realistic samples from a conditional probability distribution, and not the exact true values of a specific physical building. Notably, our two-stage workflow: using generative modeling for filling in data gaps, followed by physics-based simulation offers additional advantages in terms of preserving physical interpretability and enabling the exploration of retrofit analyses and strategies. Future work can involve expanding case studies to more complex and larger communities. 

\section{Conclusion}
This work demonstrates the potential of a diffusion-based conditional generation model that learns multivariate conditional distributions over complex building characteristics to fill gaps in building characteristics datasets.
Complete datasets with detailed
characteristics at the building-level resolution are essential in studying such fine-grained energy use patterns and developing efficient energy management strategies. Our TabDDPM-based generative approach successfully models heterogeneous building attributes in tabular datasets through a single, unified model that generates diverse, realistic values for any number of unknown characteristics conditioned on observed details. We present a comprehensive evaluation of our Conditional TabDDPM model, validating its conditional generation capabilities. This work provides a pathway to integrate generative frameworks into building energy modeling workflows, ultimately reducing the need for manual expertise in tasks such as data curation.  

\paragraph{Acknowledgements}
This work was authored by the National Laboratory of the Rockies for the U.S. Department of Energy (DOE), operated under Contract No. DE-AC36-08GO28308. This work was supported by the Laboratory Directed Research and Development (LDRD) Program at the National Laboratory of the Rockies. The views expressed in the article do not necessarily represent the views of the DOE or the U.S. Government. The U.S. Government retains and the publisher, by accepting the article for publication, acknowledges that the U.S. Government retains a nonexclusive, paid-up, irrevocable, worldwide license to publish or reproduce the published form of this work, or allow others to do so, for U.S. Government purposes.
This research was performed using computational resources sponsored by the U.S. Department of Energy's Office of Critical Minerals and Energy Innovation and located at the National Laboratory of the Rockies.






\bibliographystyle{unsrt}
\bibliography{main}
\end{document}